# Effect of a magnetic flux line on the quantum beats in the Hénon-Heiles level density

M. Brack[a], R. K. Bhaduri[b], J. Law[c], Ch. Maier[a], and M. V. N. Murthy[d]

[b] Department of Physics and Astronomy, McMaster University, Hamilton, Canada L8S 4M1

[a] Institute for Theoretical Physics, University of Regensburg, D-93040 Regensburg, Germany

[c] Department of Physics, University of Guelph, Guelph, Ontario, Canada N1G 2W1

[d] Institute of Mathematical Sciences, C.I.T. Campus, Madras 600 113, India.

## Abstract

The quantum density of states of the Hénon-Heiles potential displays a pronounced beating pattern. This has been explained by the interference of three isolated classical periodic orbits with nearby actions and periods. A singular magnetic flux line, passing through the origin, drastically alters the beats even though the classical Lagrangean equations of motion remain unchanged. Some of the changes can be easily understood in terms of the Aharonov-Bohm effect. However, we find that the standard periodic orbit theory does not reproduce the diffraction-like quantum effects on those classical orbits which intersect the singular flux line, and argue that corrections of relative order $\hbar$ are necessary to describe these effects. We also discuss the changes in the distribution of nearest-neighbour spacings in the eigenvalue spectrum, brought about by the flux line.
PACS numbers: 05.45.+b, 02.50.+s, 03.65.-w, 03.65.Ge

Typeset using REVTEX

# I. INTRODUCTION

The connection between the quantum behaviour of a particle and its classical motion in a given potential is of continuing interest. Particularly revealing is the link between the oscillating part of the quantum density of states (obtained by subtracting out its smooth, Thomas-Fermi like component) and the classical periodic orbits [1,2]. Orbits with short periods govern the large-scale energy dependence of the quantum density, while orbits with long periods determine its fine structure. It is known that even when the classical motion of a particle is globally chaotic, isolated periodic orbits may continue to exist. For that case, Gutzwiller [1] has derived the so-called "trace formula" for the oscillating part of the density of states:

$$\delta g(E) = \sum_\lambda \sum_{k=1}^\infty \mathcal{A}_{\lambda k}(E) \cos\left[k\left(\frac{1}{\hbar}S_\lambda(E) - \sigma_\lambda \frac{\pi}{2}\right)\right]. \qquad (1)$$

Here $\lambda$ goes over all primitive periodic orbits (supposed to be isolated here); $k$ counts the number of revolutions around each primitive orbit, yielding a series of harmonics; $S_\lambda(E) = \oint \mathbf{p}_\lambda \cdot d\mathbf{q}_\lambda$ is the classical action integral along the primitive orbit $\lambda$; and the Maslov index $\sigma_\lambda$ is a phase depending on its topology. The amplitudes $\mathcal{A}_{\lambda k}$ are given by

$$\mathcal{A}_{\lambda k}(E) = \frac{1}{\pi\hbar} \frac{T_\lambda}{\sqrt{|\det(\widetilde{M}_\lambda^k - I)|}} \qquad (2)$$

in terms of the period $T_\lambda = dS_\lambda/dE$ and the stability matrix $\widetilde{M}_\lambda$ of the corresponding orbit ($I$ being the 2x2 unit matrix).

If the fine structure in the quantum density of states is erased by an appropriate smoothing [2–4], then the remaining oscillations may be attributed to the lowest harmonics of the periodic orbits with the shortest periods. The objective in this paper is to study this link between the orbits of shortest period and the "coarse-grained" quantum density of states, particularly in the presence of a magnetic flux line.

The potential of Hénon and Heiles (HH) [5] is very appropriate for such a study. A classical particle can escape from this potential over three barriers if its energy is larger than the threshold energy. Below threshold the motion is confined; it becomes less regular with increasing energy and is fully chaotic at and above threshold [5]. If the particle energy is less than some 80 percent of the barrier height, then it is established [6,7] that there are only three types of distinct isolated periodic orbits of nearly equal periods and actions, all others having periods about twice as large or more. As we have shown in our earlier paper [7], the interference of these three periodic orbits can explain the beats in the quantum density of states.

In the present work, a magnetic flux line through the origin is added to the HH potential. Because of the beats in the quantum density of states, the HH potential provides a new testing ground to study the effects of a flux line and their connection to classical orbits. We find that the beat structure in the quantum density is altered drastically in the presence of the flux line. Moreover, the vortex potential which is singular at the origin introduces an additional zero in the $s$-state components of the quantum wavefunctions. The Fourier analysis of the quantum level density reveals, indeed, a new peak of approximately half the original period. In view of the fact that the classical Lagrangean equations of motion remain unaltered by the introduction of the flux line, the question arises if the changes in the quantum level density may be explained within the standard periodic orbit theory by simply adding the Aharonov-Bohm phase to the actions of those orbits that have a non-zero winding number about the origin. The answer is found to be negative.

The nearest-neighbour spacing (NNS) distributions of the quantum spectra show very interesting changes with the introduction of the flux line. The HH potential has a three-fold symmetry, and the quantum states can be classified in three distinct classes. Two of these classes have identical spectra in the absence of the flux line. This degeneracy is broken by the latter, and nontrivial changes in the NNS distributions take place. In particular, with the introduction of the flux line, the distribution for one symmetry class changes from Wigner to Poisson type – a purely quantum-mechanical effect.

Our paper is organized as follows. In section II, the quantum beats of the HH spectrum (without magnetic flux line) and their semiclassical interpretation are described. The content of this section is close to that of Ref. [7], except for a more careful calculation of the Maslov indices following the methods of Creagh, Robbins and Littlejohn [8]. Moreover, the numerically determined actions of the three leading orbits have been parametrized more accurately. We also re-analyze their degeneracy factors due to the symmetries of the HH potential, and find that Eq. (1) reproduces the quantum level density much better if one of the orbits is given an extra factor of three.

In section III, a singular flux line through the origin is introduced. We first describe the quantum results in detail and then try to interpret them in terms of the standard periodic orbit theory. A simple modification of the classical calculation, just shifting the action of the loop orbit that encloses the flux line by the Aharonov-Bohm phase, is not sufficient to fully explain the quantum results. In section IV, the NNS distributions are displayed and discussed in detail. Section V gives a short summary and the Appendix contains an account of the Maslov index calculation without the flux line.

## II. BEATS IN THE HÉNON-HEILES DENSITY OF STATES

The HH potential [5] for the planar motion of a particle is given by

$$V_{HH} = \frac{1}{2}(x^2 + y^2) + \alpha(x^2 y - \frac{1}{3}y^3). \tag{3}$$

Following the usual convention, we put $\hbar=m=\omega=1$, and $\alpha$ is a dimensionless parameter whose strength determines the anharmonicity. The equipotentials of $V_{HH}$ are shown in Fig. 1; they have a three-fold symmetry. This is obvious by writing $V_{HH}$ in polar coordinates $(r,\theta)$ in which it has the form

$$V_{HH} = \frac{1}{2}r^2 + \frac{\alpha}{3}r^3 \sin(3\theta). \tag{4}$$

Along the three symmetry axes $\theta = \pi/2$, $(\pi/2 + 2\pi/3)$, and $(\pi/2 + 4\pi/3)$, we have $\sin(3\theta) = -1$ so that the barrier height is minimum; the threshold energy there is $E^* = 1/6\alpha^2$ (in units of $\hbar\omega=1$).

We have obtained the quantum spectrum of the Hamiltonian

$$H = \frac{1}{2}(p_x^2 + p_y^2) + V_{HH} \tag{5}$$

by its diagonalization in a large harmonic oscillator basis $|nl\rangle$ with radial quantum number $n$ and angular momentum $l$ (and oscillator constant $\omega = 1$). Because of the noncentral nature of $V_{HH}$, $l$ is not a good quantum number and states with $\Delta l = \pm 3$ are mixed. This mixing takes place within three distinct and disconnected sets or symmetry classes of the basis states:

$$\{I\} : l \;\epsilon\; \{... -6, \; -3, \; 0, \; 3, \; 6...\},$$
$$\{II\} : l \;\epsilon\; \{... -5, \; -2, \; 1, \; 4, \; 7...\},$$
$$\{III\} : l \;\epsilon\; \{... -4, \; -1, \; 2, \; 5, \; 8...\}. \tag{6}$$

which together cover the full Hilbert space. Note that under time reversal, $l \to -l$, and therefore set $\{I\}$ maps onto itself, while $\{II\}$ maps onto $\{III\}$ and vice versa. In the absence of an external magnetic field, the Hamiltonian $H$ of Eq. (5) is invariant under time reversal, and thus the sets $\{II\}$ and $\{III\}$ yield identical eigenspectra. Thus $H$ may be diagonalized separately in each basis set. Another interesting point to note is that only the states belonging to the set $\{I\}$ are invariant under the rotations $\theta \to (\theta + 2\pi/3)$ and $\theta \to (\theta + 4\pi/3)$. Therefore only the states of set $\{I\}$ have the same three-fold symmetry as the Hamiltonian, but not the states belonging to the sets $\{II\}$ and $\{III\}$. This will be important for the analysis of the NNS distributions in the different spectra, which will be done separately for each symmetry class. The quantum spectrum obtained by diagonalizing $H$ is discrete even for $E > E^*$ (the barrier height) due to the truncation of the basis, and is only reliable for $E \lesssim E^*$. The basis size used in the present calculations has an energy cut-off $E_{cut} = 130 \; (\hbar\omega)$. It should be noted that in principle, even the "bound" states below the threshold have non-zero widths and the particle has a finite probability to tunnel through the barrier. The time scale relevant to our analysis is certainly much smaller than the tunneling times, so that the neglect of these widths should not affect our conclusions.

As in our earlier paper, a Strutinsky smoothing [3] is applied to obtain the overall smooth part of the density of states. Let us write the exact quantum-mechanical level density as

$$g(E) = \sum_i \delta(E - E_i), \qquad (7)$$

where the sum over $\{i\}$ includes all three sets of eigenstates. To obtain the Strutinsky smoothed density $\tilde{g}(E)$ (which is equivalent to the extended Thomas-Fermi value [9,10]), each delta function in Eq. (7) is replaced by a Gaussian of half-width $\tilde{\gamma} > 1$, multiplied by a so-called "curvature-correction" function [3] that may be written as a generalized Laguerre polynomial:

$$\tilde{g}(E) = \frac{1}{\tilde{\gamma}\sqrt{\pi}} \sum_i exp\left[\frac{(E-E_i)^2}{\tilde{\gamma}^2}\right] L_m^{(1/2)}\left[\frac{(E-E_i)^2}{\tilde{\gamma}^2}\right]. \qquad (8)$$

It can be shown [3,9] that the inclusion of this correction polynomial preserves the $2m$ lowest terms of a local Taylor expansion of the average level density. Here we use a sixth-order polynomial ($m=3$). For $\tilde{\gamma} \simeq 1.4$, all oscillations in the level density are wiped out and the so-called "plateau condition" (local independence of $\tilde{g}(E)$ of $\tilde{\gamma}$) is well fulfilled. Next, a Gaussian smoothing (without curvature-correction) is performed with $\gamma_{osc} \lesssim 0.6$, in order to obtain a "coarse-grained" quantum density of states which we denote by $g_{\gamma_{osc}}(E)$. Note that this is equivalent to suppressing all higher harmonics ($k > 1$) in the Gutzwiller trace formula (1) as well as the contributions of orbits with long time periods. Our coarse-grained oscillating density of states is thus defined as

$$\delta g(E) = [g_{\gamma_{osc}}(E) - \tilde{g}(E)]. \qquad (9)$$

The $\delta g(E)$ obtained by this procedure is shown in Figs. 2 and 3 (upper row) for three values of the anharmonicity parameter $\alpha$ and is essentially the same as that given in Ref. [7]. Note that the beating pattern for different values of $\alpha$ is approximately invariant when plotted as a function of $\alpha E$. The reason for this scaling behaviour will become clear from the semiclassical analysis below. A discrete fast Fourier transform (FFT) of $\delta g(E)$ for $\alpha = 0.04$ in the time domain clearly reveals three spikes around the period $T \approx 1$ (see Fig. 4 below). These may be identified with the periods of the three classical orbits A, B, and C that are shown in Fig. 1, averaged over the energy interval ($0 \leq E \leq 40$) used in the Fourier transform.

For obtaining the "Gutzwiller connection" to the quantum beats by means of the trace formula (1), it is essential to calculate the classical actions $S_\lambda$ of the three periodic orbits $\lambda = $ A, B, and C. This is done from the numerical solutions of the classical equations of motion. Note that the strength parameter $\alpha$ may be scaled away from the Hamiltonian $H$ in Eq. (5) by defining the scaled variables $\xi = \alpha x$ and $\eta = \alpha y$. The classical equations of motion are then universal in these coordinates with a dimensionless scaled energy $e$ defined by

$$e = E/E^* = 6\alpha^2 E. \tag{10}$$

The numerically calculated scaled actions $s_\lambda = 6\alpha^2 S_\lambda$ of the above three orbits can be parametrized, after performing a least-chi-squared fit, as

$$s_A(e) = 2\pi(e + 0.0821 e^2 + 0.0585 e^6) \qquad (0 \leq e \leq 1),$$
$$s_B(e) = 2\pi(e + 0.0698 e^2 - 0.0046 e^4) \qquad (0 \leq e \leq 2),$$
$$s_C(e) = 2\pi(e - 0.0234 e^2 + 0.0011 e^4) \qquad (0 \leq e \leq 2). \tag{11}$$

[Unlike the actions of the orbits B and C which were computed numerically, the action of orbit A can be written analytically in terms of an elliptic integral.] Rewriting the true actions $S_\lambda(E)$ as functions of the true energy variable $E$, we find from the expansions (11) the following $\alpha$ dependence:

$$S_\lambda(E, \alpha) = 2\pi[E + c_\lambda (\alpha E)^2 + \mathcal{O}(\alpha^6 E^4)] \tag{12}$$

with constant numerical factors $c_\lambda$. The first term in the action of each orbit is $\alpha$ independent and equal to the action $2\pi E$ of the harmonic oscillator, thus guaranteeing the correct $\alpha \to 0$ limit of the above fits [11]. It is the second terms, proportional to $(\alpha E)^2$, that determine the beat pattern for not too large values of $\alpha$ and $E$ [12]. This is the reason why the energy variation of the beating amplitudes of $\delta g(E)$ depends on the combination $(\alpha E)$ and thus scales with $\alpha^{-1}$, as can be observed in Fig. 2 when comparing the results for the different values of $\alpha$.

We now write down the (semi)classical expression for the oscillating part of the density of states, denoting it by $\delta g_{cl}(E)$, by just keeping the lowest harmonics $(k = 1)$ in the Gutzwiller trace formula (1):

$$\delta g_{cl}(E) = 3\mathcal{A}_A \cos\left(S_A - \sigma_A \frac{\pi}{2}\right) + 3\mathcal{A}_B \cos\left(S_B - \sigma_B \frac{\pi}{2}\right) + 2\mathcal{A}_C \cos\left(S_C - \sigma_C \frac{\pi}{2}\right). \quad (13)$$

Due to the three-fold symmetry of the HH potential, there exist three distinct and energetically degenerate orbits of type A and B, whereas the orbit C maps onto itself under the rotations about 120 and 240 degrees. Therefore, the amplitudes of A and B have been multiplied by a factor 3 in Eq. (13). The factor 2 multiplying the contribution of orbit C is to account for its two topologically different orientations, orbiting clockwise and anti-clockwise around the origin.

As is evident from Fig. 3, the agreement between the quantum and semiclassical results is much better if the contribution of orbit C is also multiplied with a degeneracy factor 3 (as has already been done in Ref. [7]). At present, we have no explanation for this fact, except for some numerical evidence from the Fourier spectrum of the quantum level density discussed in Sect. III.A below. If the orbit C had not the same discrete rotational symmetry as the HH potential, this extra factor of 3 would be justified. We find, however, no deviation from this symmetry within the numerical accuracy of our solutions, in agreement with the existing literature [6,13].

Besides the actions which we have already mentioned, the amplitudes $\mathcal{A}_\lambda$ (2) and the Maslov indices $\sigma_\lambda$ in (1) have been obtained numerically from the monodromy matrices of the three orbits (see Appendix). As already mentioned in Ref. [7], the amplitudes (2) diverge in the limit $e \to 0$, the eigenvalues of $\widetilde{M}_\lambda$ becoming equal to unity for all orbits. This happens because in the harmonic-oscillator limit $\alpha = 0$, the trace formula (1) does not apply since the periodic orbits of the harmonic oscillator are not isolated. We therefore regularize the amplitudes, imposing their exact values $2E$ [see Eq. (23) below] for the 2-dimensional harmonic oscillator by a spline fit in the low-energy region [7,12]. In Ref. [7], this was done only for the orbit B and the other amplitudes were determined through the ratio of the unsplined Gutzwiller amplitudes. We now do the splines independently for all three orbits A, B, and C.

The Maslov indices $\sigma_\lambda$ were obtained in Ref. [7] simply by counting the conjugate points of the respective orbits. As pointed out by Creagh *et al.* [8], this is strictly correct only in one dimension, while in higher dimensions there is another contribution to the Maslov index arising from the saddle-point approximation in taking the trace of the energy-dependent Green's function. We have used the methods of Creagh *et al.* [8], described in the appendix, to calculate the Maslov indices. We find that their values for the orbits A, B, and C are $\sigma_A=5$, $\sigma_B=4$, and $\sigma_C=3$, respectively. Note that these are different from Ref. [7] where their erroneous values and the slightly inaccurate parametrization of the actions for $S_B$ and $S_C$ – together with their different splines – appear to have compensated each other to give a fair agreement.

The present calculation of $\delta g_{cl}$ using Eq. (13) with the new Maslov indices yields a reasonable agreement with the quantum calculation of Eq. (9), as seen in Fig. 2 where the semiclassical results obtained according to Eq. (13) are shown in the lower row. The beat structure is reproduced with minima approximately at the correct energies. However, the interference is evidently not strong enough to give sufficient destruction at the beat minima. The disagreement is worst at the first minimum which is due the interference of orbit C with orbits A and B (having nearly identical periods, actions and amplitudes).

If we increase the amplitude of orbit C by an extra factor 3 in Eq. (13), the agreement becomes much better and now is almost perfect. The corresponding result is shown in Fig. 3. Although we cannot theoretically justify this choice, it receives strong support from the Fourier spectra discussed in the next section (see Fig. 4).

In passing, we remark that the $\alpha^{-1}$ scaling of the beating amplitudes in $\delta g(E)$, which we have explained semiclassically through Eqs. (12) and (13) above, can also be understood purely quantum-mechanically. In fact, if one treats the anharmonic part of the HH potential (3) in second-order perturbation theory, one finds for its eigenvalues the following expression [12]:

$$E_i(\alpha) = E_{nl}^{(0)} - \frac{5}{12}\left[\left(\alpha E_{nl}^{(0)}\right)^2 + \frac{7}{5}(\alpha l)^2 + \frac{11}{15}\alpha^2\right] + \ldots, \quad (14)$$

where $E_{nl}^{(0)}$ are the unperturbed harmonic-oscillator levels. Thus, if we take the $E_{nl}^{(0)}$ as the energy scale of the quantum-mechanical spectrum, we see again that the dominating $\alpha$ dependence of the approximate $E_i$ scales with $\alpha^{-1}$.

### III. EFFECTS OF THE MAGNETIC FLUX-LINE ON THE LEVEL DENSITY

We now discuss the modifications brought about by the addition of a magnetic flux line perpendicular to the $x - y$ plane at the origin of the HH potential. Its influence is most compactly expressed in the Lagrangean formalism. The Lagrangean is then given in polar coordinates by

$$L = \frac{1}{2}(\dot{r}^2 + r^2\dot{\theta}^2) - V_{HH}(r,\theta) + \delta\dot{\theta}, \quad (15)$$

where $\delta$ is the (dimensionless) strength of the flux line. The problem may also be described in terms of a particle moving in the HH potential in the presence of a magnetic field which has a delta function singularity at the origin and is zero everywhere else. The constant flux then is given by $\delta$. For $\delta = 1$, one unit of flux quantum is enclosed by any path that encloses the origin. For studying the mathematical problem, we shall assume that $\delta$ can take a continuous range of values between 0 and 1, thus allowing fractional flux quanta. This is only allowed in two-dimensional quantum mechanics [14]. From Eq. (15), it follows that the canonical momentum corresponding to the angular variable is

$$p_\theta = \frac{\partial L}{\partial \dot{\theta}} = (r^2\dot{\theta} + \delta). \quad (16)$$

Thus, the classical action $\int p_\theta d\theta$ over an orbit that encloses the origin is shifted by $\pm 2\pi\delta$ due to the existence of the flux line, according to the Aharonov-Bohm effect. However, since the extra term due to the flux line in the Lagrangean (15) is a total time derivative ($\delta$ being a constant), the *classical Lagrangean equations of motion remain unchanged*. The Hamiltonian corresponding to Eq. (15) is easily found to be

$$H = \frac{1}{2}[p_r^2 + \frac{(p_\theta - \delta)^2}{r^2}] + V_{HH}(r,\theta), \tag{17}$$

where $p_r = \dot{r}$ is the canonical radial momentum. The corresponding Schrödinger equation has to be solved numerically again, and its eigenvalue spectrum is altered non trivially. We first point out some systematics of this spectrum which will be used in analyzing both the quantum density of states and the NNS distributions. The Hamilton operator in the presence of the flux line is given by

$$\hat{H} = \frac{1}{2}[-\frac{\partial^2}{\partial r^2} - \frac{1}{r}\frac{\partial}{\partial r} + \frac{(-i\frac{\partial}{\partial \theta} - \delta)^2}{r^2}] + V_{HH}(r,\theta). \tag{18}$$

This Hamiltonian is diagonalized as before in a large harmonic oscillator basis, but this time the basis includes the flux term explicitly. This is possible because the Hamiltonian (18) with $V_{HH}$ replaced by the harmonic oscillator potential $\frac{1}{2}r^2$ can be analytically solved with eigenvalues

$$E_{nl}^{(0)}(\delta) = 2n + |l - \delta| + 1 \tag{19}$$

and the corresponding eigenvectors $|n, |l - \delta|\rangle$. Again, basis states with $\Delta l = 3$ are mixed by the nonlinear term of the HH potential. Thus, an eigenfunction has the general form

$$\Psi(r,\theta) = \overline{\sum_l} e^{il\theta} \sum_n c_{nl} R_n^{|l-\delta|}(r), \tag{20}$$

in terms of the radial parts $R_n^{|l-\delta|}(r)$ of the harmonic-oscillator basis functions. The bar over the first summation sign in Eq. (20) indicates that only the $l$ values (differing by 3 units) within one of the sets I, II or III should be summed over. The Hamiltonian (18) may be transformed to the original Hamiltonian without flux line by a gauge transformation on the wave function:

$$\Psi \rightarrow \Psi' = e^{i\delta\theta}\Psi = \overline{\sum_l} e^{i(l+\delta)\theta} \sum_n c_{nl} R_n^{|l-\delta|}(r). \tag{21}$$

Since the gauge-transformed Hamiltonian is now independent of $\delta$, the full spectrum must remain unchanged under the mapping $\delta \to (1-\delta)$, as this mapping only leads to a reshuffling of the $l$ values (recall that $0 \leq \delta \leq 1$). In particular, the spectrum at $\delta = 0$ is exactly the same as at $\delta = 1$, since all $l$ values are shifted by one unit and the global properties of the system cannot change when all states are included. Obviously, the same symmetry must be found in the level density discussed below. This has been checked numerically and found to be correct.

### A. Quantum density of states

In this subsection, the coarse-grained oscillating quantum level density $\delta g(E)$, defined in Eq. (13), will be studied for varying flux strength $\delta$. The global analysis immediately shows that it is changed drastically in the presence of the flux line. In order to make a detailed analysis, we use two windows corresponding to $\gamma_{osc} = 0.25$ and 1.0, respectively.

In Figs. 4 and 5, $\delta g(E)$ is plotted versus $E$ with $\alpha$=0.04 and 0.06, respectively, for the three values $\delta$=0, 0.25, and 0.5 of the flux strength, together with their Fourier transforms. The smooth part $\widetilde{g}(E)$ was obtained with $\widetilde{\gamma}$=1.4 as before, while $\gamma_{osc}$=0.25 was used in order to see finer details of the oscillating level density which will become evident mostly in the Fourier spectra. At $\alpha$=0.04 (Fig. 4) the Fourier spectrum is better resolved, because here we could use a larger energy range. In the absence of the flux line (top figures), the three peaks corresponding to the averaged periods of orbits A, B, and C can clearly be distinguished; C has the lowest period and B the largest. Note that the relative heights of these three peaks correspond almost quantitatively to the amplitudes $\mathcal{A}_\lambda$ (2) found numerically [cf. Fig. (4) of Ref. [7]] and averaged over the energy, if the amplitude of C is multiplied by its time-orientation degeneracy factor 2 but those of A and B are *not* multiplied by their extra symmetry degeneracy factor 3. Thus, the Fourier analysis of the quantum level density seems to support our choice to give all three orbits the same symmetry degeneracy factor which led to Fig. 3.

We now consider the effects of the flux line on the beats in Figs. 4 and 5. Notice that only one beat minimum persists at all values of $\delta$ that are shown. This minimum is essentially due to the interference of orbits A and B. In the special case $\delta=0.25$ it is the only minimum present in the energy range considered, and the peak corresponding to orbit C is missing in the Fourier spectrum. A simple explanation of this fact will be offered in the next subsection. The peak belonging to orbit C appears again at $\delta=0.5$ along with some of the original beat structure.

A very interesting phenomenon appears when $\delta g(E)$ is examined with a stronger coarse-graining, obtained with $\gamma_{osc}=1.0$, that corresponds to an emphasis on periods shorter than the fundamental period $T \simeq 1$. Since the periods of the primitive orbits A, B, and C are all close to one (at least at $\delta=0$), we should expect all oscillations of the level density to die out. This is, indeed, the case as seen on the top row of Fig. 6. However, for $0 < \delta < 1$, $\delta g(E)$ shows a remarkably neat oscillatory behaviour with almost no visible beat pattern. The corresponding Fourier spectra are also shown in Fig. 6 and exhibit a clear structure around the period $T \approx 1/2$ (in units of $2\pi/\omega$) in the time domain. The origin of this "frequency doubling" in the quantum-mechanical framework is clearly the presence of the singular vortex interaction at the origin. When $\delta=0$, the $s$-state ($l = 0$) components of the wavefunctions would be non-zero at the origin. But when $\delta \neq 0$, a genuine zero in the wavefunction is introduced at the origin. The frequency doubling may thus be interpreted as a diffraction effect of an incoming wave at the location of the flux line: the transmitted wave corresponds to the full period, whereas the reflected wave gives rise to a Fourier component with a fraction of this period. Indeed, Fig. 6 seems to indicate two peaks around $T \approx 0.5$. In anticipation of the semiclassical interpretation below, we emphasize that this is a pure quantum effect, since a classical particle with zero angular momentum cannot penetrate the flux line. Also the fact that the quantum-mechanical spectrum is unaltered by the flux line when its strength is $\delta=1$ can not be understood classically.

The scattering of a free charged particle by a singular flux line has been studied in the classic paper by Aharonov and Bohm [15]. These authors have shown that the scattering cross section goes like $\sin^2(\pi\delta)/\cos^2(\theta/2)$ (for $\theta \neq \pi$), where $\theta$ is the scattering angle. It vanishes for both $\delta=0$ and $\delta=1$. Although the particles in the HH potential are not free waves, the features discussed above bear some resemblance to the result of Aharonov and Bohm. In particular, the amplitude of the reflected wave ($\theta=0$) having a maximum at $\delta=0.5$ seems to agree qualitatively with the amplitude of the frequency-doubled peak seen in the Fourier spectrum of Fig. 6. Unfortunately, the above expression for the scattering cross section does not hold for $\theta=\pi$, so that no definite conclusions can be drawn for the amplitude of the transmitted wave.

To illustrate this frequency doubling and its genuineness as a quantum phenomenon, consider the simple example of particle in a spherical two-dimensional harmonic oscillator potential with a flux line at the origin. (This corresponds to choosing $\alpha=0$ in the HH potential.) The exact level density can be computed analytically from the spectrum $E_{nl}^{(0)}(\delta)$ given in Eq. (19) and is given by

$$g(E,\delta) = \frac{1}{(\hbar\omega)^2}\left[E\left\{1 + 2\sum_{k=1}^{\infty}\cos(2\pi k\delta)\cos(2\pi kE/\hbar\omega)\right\}\right.$$
$$\left. - 2\delta\hbar\omega\sum_{k=1}^{\infty}\sin(2\pi k\delta)\sin(2\pi kE/\hbar\omega) + \hbar\omega\sum_{k=1}^{\infty}(-1)^k\sin(\pi k\delta)\sin(2\pi kE/2\hbar\omega)\right], \quad (22)$$

where we have put in the factors $\hbar\omega$ explicitly. Notice that in the limits $\delta=0$ and $\delta=1$, only the first term remains that gives the exact level density of the unperturbed harmonic oscillator:

$$g(E) = \frac{E}{(\hbar\omega)^2}\left\{1 + 2\sum_{k=1}^{\infty}\cos(2\pi kE/\hbar\omega)\right\}. \quad (23)$$

Note that the oscillatory terms are of the form $\cos(kS/\hbar)$ with $S(E) = 2\pi E/\omega$. The period of the fundamental is $T_0 = dS/dE = 2\pi/\omega$. In Eq. (22), the corrections due to the flux line have an amplitude of order $\hbar$ relative to the leading term. In particular, in the second correction term we see that the lowest harmonic has the frequency $2\omega$, i.e. a period $T_0/2$. This corresponds to the wave reflected at the singularity, while the transmitted component is contained in the first two terms which have the same fundamental period $T_0$ as the unperturbed oscillator.

The fact that the new frequency-doubled components appear as $\hbar$ corrections to the leading-order terms introduces a *caveat* for the naive interpretation of these results in terms of classical periodic orbits. In order to include diffraction effects, such as those brought about by the singular flux line, it is evidently necessary to introduce quantum corrections to the standard periodic orbit theory.

### B. Semiclassical analysis and its limitations

In our semiclassical analysis at $\delta=0$ in Sect. 2 we found a reasonable interpretation of the quantum-mechanical level density in terms of the three isolated periodic orbits with shortest periods. Now we attempt a semiclassical interpretation of the case with the flux line turned on. As emphasized above, we cannot expect to classically simulate the quantum effects of diffraction, but we we can still explain some of the observed features in the level density and the Fourier spectrum at the leading order. For example, consider the special case with $\delta=0.25$. As we see in Fig. 4, the contribution from orbit C seems to disappear at this flux strength. To understand this, recall that, unlike A or B, the loop orbit C encloses the flux line and therefore its action is shifted by $\pm 2\pi\delta$, as already stated above. For $\delta = 0.25$, this shift is $\pm\pi/2$ where the plus and minus signs are associated with clockwise and anti-clockwise motion of the particle. The total contribution of the orbit C to $\delta g(E)$, assuming the amplitude $\mathcal{A}_C$ not to depend on the sign of $\delta$, is then found to vanish:

$$\mathcal{A}_C[\cos(S_C^0 - \sigma_c\pi/2 + \pi/2) + \cos(S_C^0 - \sigma_c\pi/2 - \pi/2)] = 0. \tag{24}$$

In the above equation, $S_C^0$ is the action for the orbit C for $\delta = 0$, and $\sigma_c$ is the Maslov index whose precise value is irrelevant for this argument. This destructive interference between the two time-reversed orbits C by the Aharonov-Bohm phase is, of course, also a quantum phenomenon, but it can be incorporated at the leading order of the periodic orbit theory. However, at $\delta=0.25$, we still see in Fig. 4 the two periods in the Fourier spectrum corresponding to the periods of A and B. Since these orbits do not enclose the flux line, we may assume that they remain unaffected. However, if we now use Eq. (13) and omit the contribution from orbit C, the resulting semiclassical $\delta g(E)$ reproduces the beat minimum seen in Figs. 4 and 5 for $\delta=0.25$ at approximately the right energy, but fails to give the correct form of the envelope. This should be clear from the fact that the orbit A required for this beat does not exist classically, but is mimicked quantum-mechanically by the transmitted wave. The failure becomes even more dramatic at $\delta=0.5$, as shown in Fig. 7, if we use Eq. (13) with the unaffected Gutzwiller amplitudes and Maslov indices obtained at $\delta=0$, but add the Aharonov-Bohm phase for the orbit C:

$$\begin{aligned}\delta g_{cl} &= 3\mathcal{A}_A \cos\left(S_A^0 - 5\pi/2\right) + 3\mathcal{A}_B \cos\left(S_B^0 - 2\pi\right) \\ &\quad + \mathcal{A}_C \left[\cos\left(S_C^0 - 3\pi/2 + 2\pi\delta\right) + \cos\left(S_C^0 - 3\pi/2 - 2\pi\delta\right)\right] \\ &= 3\mathcal{A}_A \cos\left(S_A^0 - 5\pi/2\right) + 3\mathcal{A}_B \cos\left(S_B^0 - 2\pi\right) - 2\mathcal{A}_C \cos\left(S_C^0 - 3\pi/2\right). \quad (\delta = 1/2) \quad (25)\end{aligned}$$

This limitation of the standard semiclassical analysis calls for a systematic investigation of quantum corrections to the leading-order terms provided by the periodic classical orbits. Such corrections might arise either from extensions of the saddle-point approximation, or from corrections going beyond the leading order which would involve new, not necessarily closed orbits.

## IV. EFFECT OF THE MAGNETIC FLUX LINE ON THE NNS DISTRIBUTIONS

While a study of the quantum density of states of a classically chaotic system unravels its global behaviour, the local fluctuations are often analyzed by studying the nearest neighbour spacing (NNS) distributions in the eigenspectrum [16]. In general the NNS distribution is expected to be of Poisson type for integrable systems while it is of Wigner type for classically chaotic systems, though exceptions to this rule are known [17]. In this section the NNS distributions for the HH system are analyzed with and without the flux line.

Recall that the HH Hamiltonian is block diagonal and the eigenvalues can be put in three distinct and disconnected sets, namely I,II and III (see section I) due to the three-fold symmetry of the Hamiltonian. The NNS distributions are therefore analyzed in each of these classes separately. The symmetry classes are preserved even in the presence of the flux line, since the term corresponding to the coupling of the particle to the flux line is separately invariant under rotations. The NNS, for a sequence of eigenvalues $\{E_1, E_2, ...\}$ is defined through the following unfolding procedure [17]

$$s_i = (E_{i+1} - E_i)g(E_i), \qquad (26)$$

where $g(E_i)$ is the full density of states of the system evaluated locally and $s_i$ is the dimensionless nearest neighbour spacing. The sequence of the $s_i$ thus generated are put into various bins to generate the NNS distribution denoted by $P(s)$. Typically about 300 or more states are included in the analysis to obtain sufficient statistics. This restricts the range of the HH parameter to be $\alpha \leq 0.08$, since the number of bound states below the barrier energy decreases with increasing $\alpha$.

We first consider the case $\delta=0$ where there is no flux line. We restrict here the discussion to the case $\alpha = 0.06$, having assured that one finds the expected signatures for the transition from regular to chaotic behaviour as $\alpha$ is increased from 0.04 to 0.08. For $\alpha = 0.06$, we have typically about 400 eigenvalues below the barrier energy $E^*$=46.3 and we scan the energy range $16 \leq E \leq 43$, omitting the low-energy end. The $P(s)$ distributions are shown separately for the sets I, II and III in Figs. 8, 9, and 10, respectively. At $\delta$=0, the $P(s)$ distributions for Set II (Fig. 9a) and Set III (Fig. 10a) are identical and have more resemblance to Wigner-type distributions than the Poisson type, while for Set I (Fig. 8a) they are more of Poisson type. This may be understood, as mentioned earlier, from the fact that all the states in Set I have the same three-fold symmetry of the HH Hamiltonian since $l$ is always a multiple of 3. This symmetry is lost in the sets II and III which, however, have identical eigenvalues and hence identical $P(s)$ distributions. A similar effect is found in the case of a particle moving in a rectangular enclosure in the presence of a flux line, where the $P(s)$ distribution for states which share the symmetry of the Hamiltonian (in this case the parity) is of Poisson type, whereas for other sets it is more like a Wigner distribution [18].

In the presence of the flux line, some nontrivial changes are introduced in the NNS distribution. We first summarize the changes and then put forward the reasons for the same. The effect of the flux line on the NNS distributions is shown in Figs. 8 – 10 for $\delta =0$, 0.231, 0.333, 0.5, 0.769, and 1.0, labeled a,b,...,f within each figure. It is obvious that there are some nontrivial features. Even though the $P(s)$ distributions are close to either Poisson of Wigner for $\delta = 0$ and $\delta = 1$ they do not have these generic feature for arbitrary values of $\delta$. However, notice that the $P(s)$ distributions for the set I and set II are related (see Figs. 8 and 9),

$$P_I(\delta) \Longleftrightarrow P_{II}(1-\delta) \qquad (27)$$

and also (see Fig. 10)

$$P_{III}(\delta) \Longleftrightarrow P_{III}(1-\delta), \qquad (28)$$

where the subscript denotes the set of eigenvalues. At the outset the mapping $\delta$ to $(1-\delta)$ appears peculiar, but actually there is a hint on the analytic structure of the eigenvalues even though its full form cannot be obtained analytically. From Eq. (20) it is obvious that the eigenvalues in each set in principle depend on all $l$'s in that given set. Eq. (21) also indicates that these $l$'s are all shifted by $\delta$ in the presence of the flux line. However we still do not know how the eigenvalues depend on $l$ and $\delta$. The above relations between $P(s)$ distributions in each set suggest that the eigenvalues can only be functions of the form $f(|l_i - \delta|)$ with the dependence of $l_i$ and $\delta$ restricted to be of the form, $|l_i - \delta|$. To see how this works, consider the mapping $\delta \to (1-\delta)$, then

$$|l_i - \delta| \to |l_i - (1-\delta)| = |-(l_i - 1) - \delta|. \qquad (29)$$

That is a shift in $\delta \to (1-\delta)$ is equivalent to $l_i \to -(l_i - 1)$ and therefore

$$\begin{aligned} \text{set I: } \{...,-6,-3,0,3,6,...\} &\leftrightarrow \text{set II: } \{...,7,4,1,-2,-5,...\}, \\ \text{set III: } \{...,-7,-4,-1,2,5,...\} &\leftrightarrow \text{set III: } \{...,8,5,2,-1,-4,...\}, \end{aligned} \qquad (30)$$

which explains the observed results for $\delta \to (1 - \delta)$. In the absence of the cubic term in $V_{HH}$, it is known analytically that the eigenvalues depend on $l$ and $\delta$ through $|l - \delta|$. It is therefore interesting that this form carries over to the full HH potential with the flux line.

It may also be noticed that the envelope of the $P(s)$ distribution changes from Poisson to almost Wigner shape in set I as a function of $\delta$, while the opposite is the case for set II. This quantum effect may be understood by noting that as $\delta$ is increased, set II goes over to set I, thereby introducing more symmetry (see the discussion after Eq. 5). Although $P(s)$ distributions for a billiard in the presence of a singular flux line have been studied before [19], their interpolation property with the fractional strength of the flux line is novel.

## V. SUMMARY

We have interpreted the coarse-grained quantum level density of the Hénon-Heiles potential in terms of the periodic orbit theory. The pronounced beats can be explained by the interference of three types of short periodic orbits A, B, and C which have nearly identical actions and periods in the energy region well below the barriers. There is, however, a problem of counting the orbits. The conventional counting of different isolated primitive orbits, taken together with their numerically determined amplitudes in the Gutzwiller trace formula, does not seem to agree with the amplitudes found in the Fourier transform of the quantum density, unless an extra weight of 3 is given to the orbit C. Including this extra factor, the lowest harmonics of the trace formula give an almost quantitative agreement with the coarse-grained quantum density.

In our future studies, we plan to investigate the counting problem in more detail by studying other potentials with isolated periodic orbits.

The inclusion of a magnetic flux line through the origin causes substantial changes in the quantum spectrum, affecting the beats in the level density dramatically. A straight-forward application of the trace formula, adding the Aharonov-Bohm phase to the action of the loop orbit C that surrounds the flux line, fails to yield a satisfactory agreement with the quantum level density (both with and without the extra factor 3 for orbit C), although the disappearance of the Fourier peak corresponding to orbit C at the flux strength $\delta = 0.25$ can be understood in terms of the Aharonov-Bohm phase. Additionally, quantum scattering of the particle by the singular flux line takes place. This creates a new peak in the Fourier spectrum that corresponds to a reflected wave, simulating a new classical orbit of about half the period of the other leading orbits, while the Fourier signal of the transmitted wave – whose corresponding classical full period should no longer exist – still persists. To emphasize this point, the analytical expression of the quantum level density for a harmonic oscillator potential with a flux line through the origin is presented. It shows that corrections of order $\hbar$ with respect to the leading terms arise in the level density when the flux line is turned on.

It is likely that the calculation of Maslov indices has to be modified in the presence of the vector potential that describes the flux line, in order to take account of the difference between the mechanical and the canonical momenta. The determination of higher-order quantum corrections to the Gutzwiller formula, in the presence of a flux line, poses a challenging problem for future research.

Although no major surprises occurred in studying the nearest-neighbour-spacing distributions of the Hénon-Heiles spectrum, they were found to exhibit novel interpolation properties as the flux strength is varied from zero to unity.

We are grateful to Stephen Creagh for his invaluable help in understanding and computing the Maslov indices. Thanks are also due to J. Lefebvre, A. Magner, H. Nishioka and S. Reimann for stimulating discussions. This research was supported by the Natural Sciences and Engineering Research Council of Canada. M.B. and M.V.N.M. acknowledge the hospitality of the Department of Physics and Astronomy at McMaster, and R.K.B. that of the Institute of Mathematical Sciences at Madras. M.V.N.M. acknowledges the financial support from the Hans-Vielberth-Universitätsstiftung Regensburg.

# VI. APPENDIX: MONODROMY MATRIX AND CALCULATION OF THE MASLOV INDICES

For the calculation of the Maslov indices of the three orbits A, B, and C, we have followed the methods given by Creagh *et al.* [8]. In general, the overall index $\sigma$ appearing in the trace formula (1) for a given periodic orbit is a sum of two contributions:

$$\sigma = \mu + \nu, \tag{31}$$

whereof the first part $\mu$ is the Maslov index occurring in the semiclassical expression of the Green's function $G(\mathbf{r}, \mathbf{r}'; E)$ [1] and counts the number of conjugate points of a given orbit at fixed energy. The second contribution $\nu$ arises when taking the trace of $G(\mathbf{r}, \mathbf{r}'; E)$ in order to arrive at the level density $g(E)$ (or its oscillating part). Whereas both $\mu$ and $\nu$ may depend on the starting point along a periodic orbit, their sum has been shown [8] to be a topological invariant.

Two different procedures are used, depending on whether an orbit is stable or unstable. We shall sketch these methods very briefly below and refer to Creagh *et al.* for their derivation. But first we recall the definition of the monodromy matrix.

We start from a periodic orbit given in terms of the coordinates $\mathbf{q} = (x, y)$ and momenta $\mathbf{p} = (p_x, p_y)$ as functions of time: $\mathbf{q}(t) = \mathbf{q}(t + T)$, $\mathbf{p}(t) = \mathbf{p}(t + T)$. The stability of the orbit is given by the propagation of small perturbations $\delta\mathbf{q}(t)$, $\delta\mathbf{p}(t)$ away from the exact solution $\mathbf{q}_0(t)$, $\mathbf{p}_0(t)$:

$$\mathbf{q(t)} = \mathbf{q}_0(t) + \delta\mathbf{q(t)}; \qquad \mathbf{p(t)} = \mathbf{p}_0(t) + \delta\mathbf{p(t)}. \tag{32}$$

These perturbations satisfy the linearized equations of motion

$$\frac{d}{dt}\begin{pmatrix} \delta\mathbf{q(t)} \\ \delta\mathbf{p(t)} \end{pmatrix} = \begin{pmatrix} \mathbf{0} & \mathbf{1} \\ -\mathbf{V}_0(t) & \mathbf{0} \end{pmatrix} \begin{pmatrix} \delta\mathbf{q(t)} \\ \delta\mathbf{p(t)} \end{pmatrix}, \tag{33}$$

with $T$-periodic coefficients given by the second derivatives of the potential $V(\mathbf{q})$ along the orbit:

$$\mathbf{V}_0(t) = \mathbf{V}_0(t+T) = V_{ij}(\mathbf{q}_0(t)) = \left. \frac{\partial^2 V(\mathbf{q})}{\partial q_i \partial q_j} \right|_{\mathbf{q}=\mathbf{q}_0(t)}. \tag{34}$$

The time evolution of the solutions of the linear differential equations (33) is given by the matrizant $X(t)$:

$$\begin{pmatrix} \delta\mathbf{q}(t) \\ \delta\mathbf{p}(t) \end{pmatrix} = X(t) \begin{pmatrix} \delta\mathbf{q}(0) \\ \delta\mathbf{p}(0) \end{pmatrix} \tag{35}$$

with the initial condition $X(0)=\mathbf{1}$. The value of $X(t)$ after one period $T$ is called the monodromy matrix $M$:

$$M = X(T). \tag{36}$$

According to the Lyapounov theorem, the (4x4) monodromy matrix $M$ has two pairs of eigenvalues which are mutually inverse. For a conservative system, two of them are unity, corresponding to small perturbations along the orbit. We can therefore, after a transformation to the intrinsic coordinate system of the orbit, write $M$ in the form

$$M = \begin{pmatrix} \widetilde{M} & \mathbf{0} \\ \mathbf{0} & \begin{pmatrix} 1 & 0 \\ 0 & 1 \end{pmatrix} \end{pmatrix}; \tag{37}$$

where $\widetilde{M}$ is the (2x2) stability matrix. If the orbit is stable, its eigenvalues lie on the unit circle with values $e^{\pm i\chi}$ and $0 \leq \chi \leq \pi$. For unstable orbits, the eigenvalues are real of the form $\pm e^{\pm \chi}$, $\chi > 0$ being the Lyapounov exponent. In our numerical calculations, we have obtained $M$ using the monodromy matrix method of Baranger *et al.* [22].

**a) Unstable orbits:**

For an unstable orbit it is possible to obtain the total Maslov index $\sigma$ at once from the full monodromy matrix $M$. For that purpose one chooses, at a suitable starting point of the orbit (at time $t = 0$), the eigenvector $\mathbf{e}_0 = (e_x, e_y, e_{p_x}, e_{p_y})$ that corresponds to the unstable eigenvalue of $M$, and the flow vector of the Hamiltonian, $\mathbf{f}_0 = (\dot{x}, \dot{y}, \dot{p}_x, \dot{p}_y)$. One then propagates these two vectors once around the orbit using the matrizant $X(t)$:

$$\mathbf{e}(t) = X(t)\,\mathbf{e}_0, \qquad \mathbf{f}(t) = X(t)\,\mathbf{f}_0, \tag{38}$$

which yields two time dependent vectors $\mathbf{e}(t) = e_i(t)$ and $\mathbf{f}(t) = f_i(t)$ ($i = 1,2,3,4$). Next one defines two real (2x2) matrices $U(t)$ and $V(t)$ by

$$U(t) = \begin{pmatrix} e_1(t) & f_1(t) \\ e_2(t) & f_2(t) \end{pmatrix} \qquad V(t) = \begin{pmatrix} e_3(t) & f_3(t) \\ e_4(t) & f_4(t) \end{pmatrix}. \tag{39}$$

The Maslov index $\sigma$ is then found as the winding number of the complex quantity $C(t) = \{\det[U(t) - iV(t)]\}^2$, following its path in the complex plane during one full period $T$. Fig. 11 shows this path for orbit A at the scaled energy e=0.9 where it is unstable; clearly the winding number is $\sigma_A = 5$.

**b) Stable orbits:**

For stable orbits it is not easy to single out a vector that is orthogonal to the orbit, and the two contributions to $\sigma$ in Eq. (31) have to be computed separately.

The calculation of $\mu$ is obtained by a similar procedure as that outlined above for $\sigma$ of an unstable orbit. One takes the flow vector $\mathbf{f}_0$ and combines it with the vector $\mathbf{e}_0 = (0,0,1,-\dot{x}/\dot{y})$, assuming that $\dot{y} \neq 0$ at the starting point or, equivalently, $\mathbf{e}_0 = (0,0,-\dot{y}/\dot{x},1)$, if $\dot{x} \neq 0$ at the starting point. One propagates these two vectors along the orbit using Eq. (38) and defines $U$ according to Eq. (39). The Maslov index $\mu$ is then found as the number of zeros which det$U$ assumes during one full period, hereby not counting the zero that appears trivially at the starting point.

The index $\nu$ is given by the sign of the quantity $w$ defined by

$$w = \frac{\mathrm{Tr}\widetilde{M} - 2}{b}, \tag{40}$$

where $b$ is the upper right element of the stability matrix:

$$\widetilde{M} = \begin{pmatrix} a & b \\ c & d \end{pmatrix}, \tag{41}$$

which is also given by

$$b = \frac{dq_\perp(t=T)}{dp_\perp(t=0)}. \tag{42}$$

Hereby, $p_\perp(t)$ and $q_\perp(t)$ are the momentum and coordinate, respectively, perpendicular to the orbit. If $w$ is positive, we have $\nu=0$ and if $w$ is negative, then $\nu=1$. The calculation of $b$ (42) in principle requires a transformation to the intrinsic coordinate system of the orbit and is not simple in general. However, for specific cases, such as the stable orbits A and C in the HH potential, the sign of $b$ can be found relatively easily numerically, when solving the equations of motion on a computer and following the time evolution of a slightly perturbed periodic orbit on the screen: one gives it a small perpendicular starting velocity at time $t_0$ and determines the sign of the perpendicular coordinate after one revolution around the perturbed orbit, i.e. at time $t_0 + T$.

Figure 12 illustrates these methods of determining $\mu$ and $\nu$ for orbit A at the scaled energy $e=0.8$ where it is still stable.

Both above recipes for obtaining $\mu$ and $\nu$ can also be used for unstable orbits; we have done this as a check of the numerical methods and found the sum $\mu + \nu$ always to agree with the $\sigma$ found as a winding number as described above. Note that in all the above computations of $\sigma$, $\mu$ and $\nu$, one may *not* start at any turning point of a given periodic orbit since singularities or discontinuities can occur at turning points in some of the calculated quantities. Thus, for the orbits A we started at the origin, and for the orbits B we started at the apex (intersection with a symmetry axis of the HH potential).

For the special case of a stable isolated periodic orbit without turning points, Gutzwiller [1] has shown that, by a somewhat different bookkeeping of the phases, an alternative form of the trace formula can be given in which the total Maslov index cancels. The contribution of this orbit to $\delta g(E)$ is then

$$-\frac{1}{\hbar\pi} \sum_{k=1}^{\infty} \frac{T}{2\sin(k\chi/2)} \sin(kS/\hbar), \tag{43}$$

if the eigenvalues of $\widetilde{M}$ are $e^{\pm i\chi}$ with $0 < \chi \leq \pi$. It is easy to see that for orbit C with the Maslov index $\sigma_C=3$ this leads, indeed, to the same contribution to $\delta g(E)$ as that found from Eqs. (1) and (2).

# REFERENCES

[1] M. C. Gutzwiller: *Chaos in Classical and Quantum Mechanics* (Springer-Verlag, New York, 1990); J. Math. Phys. **10**, 1004 (1969); *ibid.*, **11**, 1791 (1970); *ibid.*, **12**, 343 (1971).

[2] R. Balian and C. Bloch, Ann. Phys. (NY) **69**, 76 (1972).

[3] V. M. Strutinsky, Yad. Fiz. **3**, 614 (1966); Nucl. Phys. **A 95**, 420 (1967); *ibid.*, **A 122**, 1 (1968).

[4] M. Sieber and F. Steiner, Phys. Lett. **A 144**, 159 (1990).

[5] M. Hénon and C. Heiles, Astr. J. **69**, 73 (1964).

[6] K. T. R. Davies, T.E. Huston, and M. Baranger, Chaos **2**, 215 (1992).

[7] M. Brack, R. K. Bhaduri, J. Law and M. V. N. Murthy, Phys. Rev. Lett. **70**, 568 (1993).

[8] S. C. Creagh, J. M. Robbins and R. G. Littlejohn, Phys. Rev. **A42**, 1907 (1990).

[9] M. Brack and H. C. Pauli, Nucl. Phys. **A 207**, 401 (1973).

[10] R. K. Bhaduri and C. K. Ross, Phys. Rev. Lett. **27**, 606 (1971); B. K. Jennings, Ann. Phys. (NY) **84**, 1 (1974).

[11] Note that the fits given in Ref. [7], although numerically valid in the energy regions given, did not respect this harmonic-oscillator limit.

[12] Ch. Maier, Thesis, University of Regensburg (1993, unpublished).

[13] R. C. Churchill, G. Pecelli, and D. L. Rod, in *Stochastic Behaviour in Classical and Quantum Hamiltonian Systems*, Eds. G. Casati and J. Ford (Springer-Verlag, N.Y., 1979) p. 76.

[14] R. Jackiw, Ann. Phys. **201**, 83 (1990); A. S. Goldhaber and R. Mackenzie, Phys. Lett. **214B**, 471 (1990).

[15] Aharonov and D. Bohm, Phys. Rev. **115**, 485 (1959).

[16] S. W. McDonald and A. N. Kaufman, Phys. Rev. Lett. **42**, 1189 (1979).

[17] Hua Wu, M. Vallières, D. H. Feng, and D. W. L. Sprung, Phys. Rev. **A42**, 1027 (1990).

[18] G. Date, S. R. Jain and M. V. N. Murthy, Preprint IMSc/93-47.

[19] M. V. Berry and M. Rubnik, J. Phys. A: Math. Gen. **19**, 649 (1986); M. Rubnik and M. V. Berry, J. Phys. A: Math. Gen. **19**, 669 (1986).

[20] J. M. Leinaas and J. Myrheim, Nuovo Cim. **370**, 1 (1977).

[21] S. Reimann, M. Brack, A. Magner and M. V. N. Murthy, to be published.

[22] M. Baranger, K. T. R. Davies, and J. H. Mahoney, Ann. Phys. (N.Y.) **186**, 95 (1988).

# FIGURES

FIG. 1. Equipotential lines of the Hénon-Heiles potential in the plane $\xi, \eta$ of scaled coordinates. The straight lines correspond to the threshold energy $E = E^*$. The three classical periodic orbits of shortest lengths A ("linear"), B ("smiley") and C ("loop") for $E = E^*$ (i.e., $e = 1$) are displayed by the heavy lines.

FIG. 2. Oscillating part of the level density of the Hénon-Heiles Hamiltonian, for the values $\alpha$=0.04 (left), 0.06 (middle) and 0.08 (right). *Upper row:* Coarse-grained quantum-mechanical results defined by Eq. (9). *Lower row:* Result of the classical periodic orbit calculation, Eq. (13).

FIG. 3. Same as Fig. 2, but in the lower row, the contribution of orbit C to Eq. (13) has been multiplied by 3.

FIG. 4. *Left column:* Quantum-mechanical level density $\delta g(E)$ in the presence of a flux line with strength $\delta$=0.0 (top), 0.25 (middle), and 0.5 (bottom) for the HH potential with $\alpha$=0.04. The coarse-graining was done with a smoothing width $\gamma_{osc} = 0.25$. *Right column:* Fourier Transform of the $\delta g(E)$ shown in the left column. The $x$-axis gives the time period $T$ in units of $2\pi/\omega$.

FIG. 5. Same as Fig. 4, for $\alpha = 0.06$.

FIG. 6. Same as Fig. 4, for $\alpha = 0.06$ and $\gamma_{osc} = 1.0$.

FIG. 7. Oscillating part of the HH level density for $\alpha$=0.04 without flux line (left) and with flux line with strength $\delta = 0.5$ (right). *Top row:* Quantum-mechanical result. *Middle row:* Semiclassical result Eq. (25). *Bottom row:* Same as middle row, but multiplying the contribution of orbit C in Eq. (25) by 3.

FIG. 8. $P(s)$ distributions for the eigenvalue set I for $\delta = 0, 0.231, 0.333, 0.5, 0.769$, and $1.0$ at $\alpha = 0.06$. The dashed line shows a Wigner distribution and the dotted line a Poisson distribution.

FIG. 9. Same as Figure 8 for the eigenvalue set II.

FIG. 10. Same as Figure 8 for the eigenvalue set III.

FIG. 11. Illustration of the Maslov index calculation for the unstable orbit A at scaled energy $e$=0.9. Plotted is the complex quantity $C(t)$ (see text) in the complex plane over one period. (Start and end points are indicated by arrows.) The winding number is 5, giving $\sigma_A = 5$.

FIG. 12. Illustration of the Maslov index calculation for the stable orbit A at scaled energy $e$=0.8, having $\sigma_A = 5$. *Upper part:* One period of the slightly perturbed orbit A. (The unperturbed orbit A, a straight line at polar angle 30 degrees, is shown by the thin dashed line.) Here we start at the origin $\xi=\eta=0$ at an angle of 32 degrees, thus with a positive perpendicular velocity $p_\perp(0)$. After one period $T$, the perturbed orbit (shown by the heavy line) ends with a positive perpendicular coordinate $q_\perp(T)$, giving $b > 0$ and thus $\nu$=1 (since $\mathrm{Tr}\widetilde{M}_A < 2$; see text). *Lower part:* The quantity $\det U$ (see text) as a function of time $t$ over one full period, exhibiting four zeros (with $t > t_0$) and thus giving $\mu$=4.

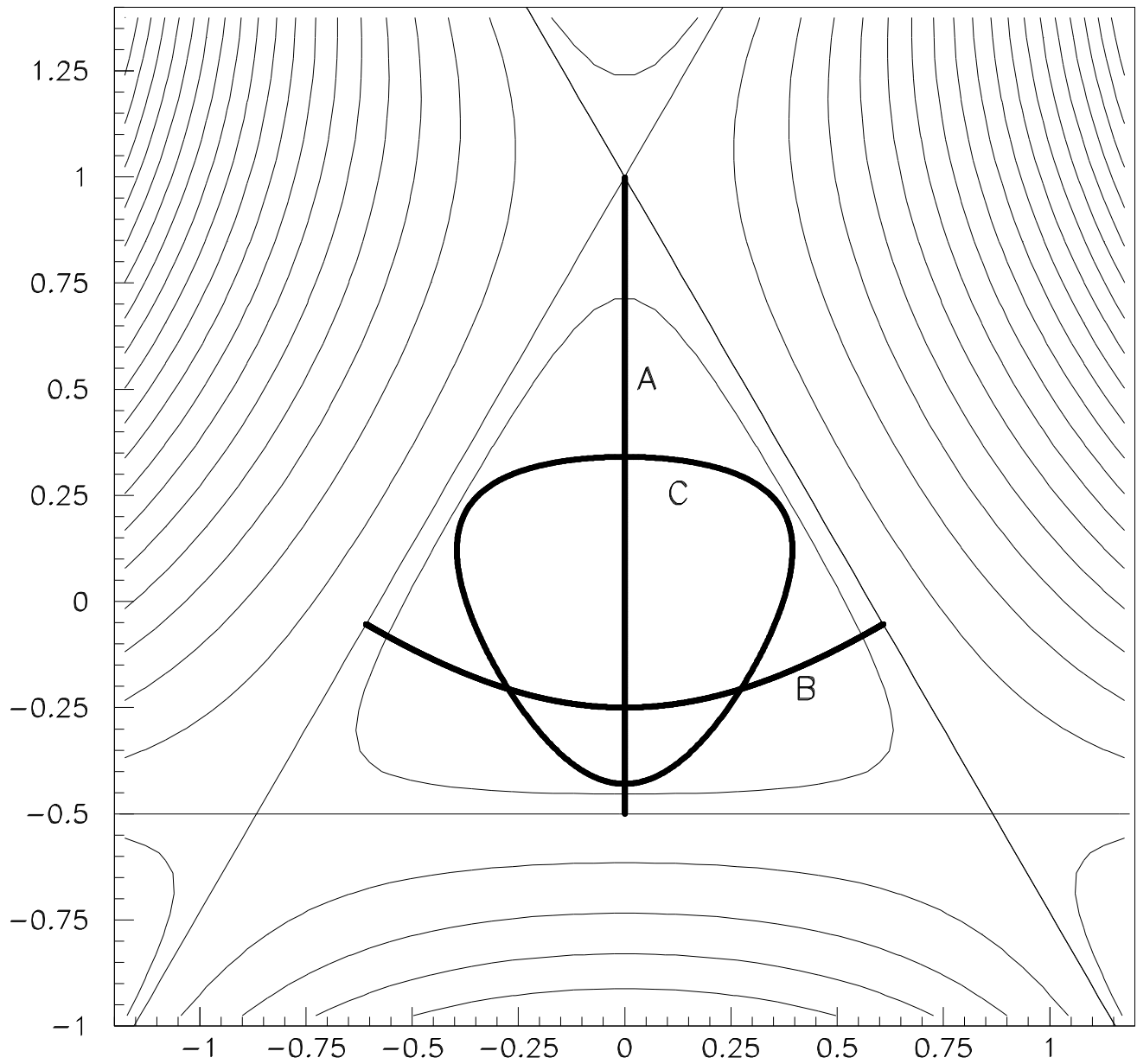

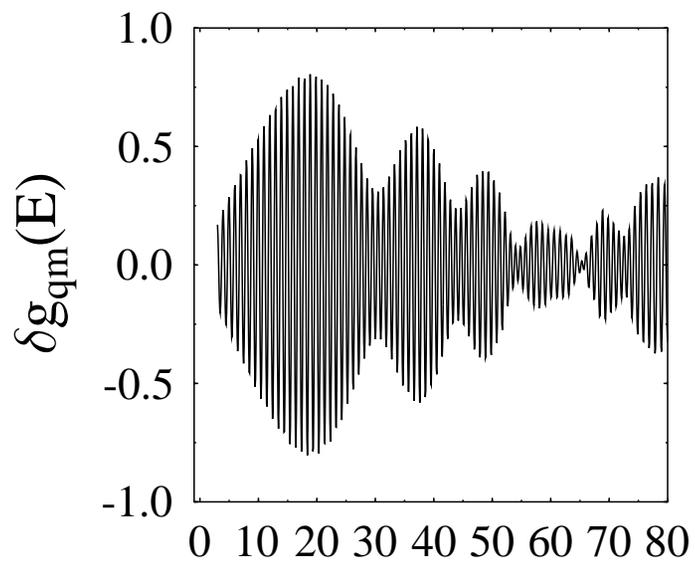 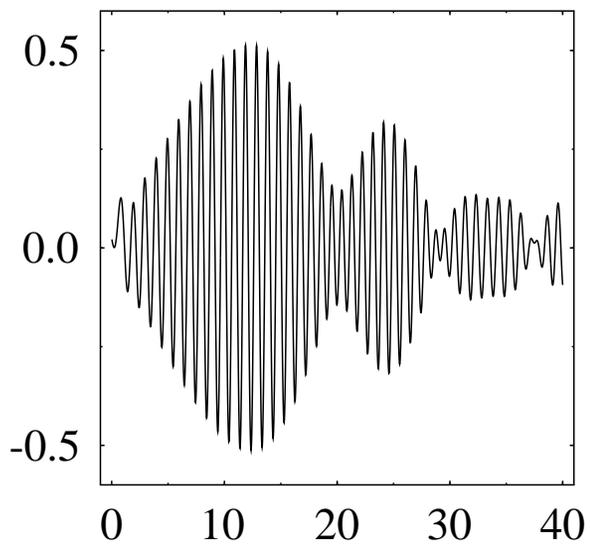 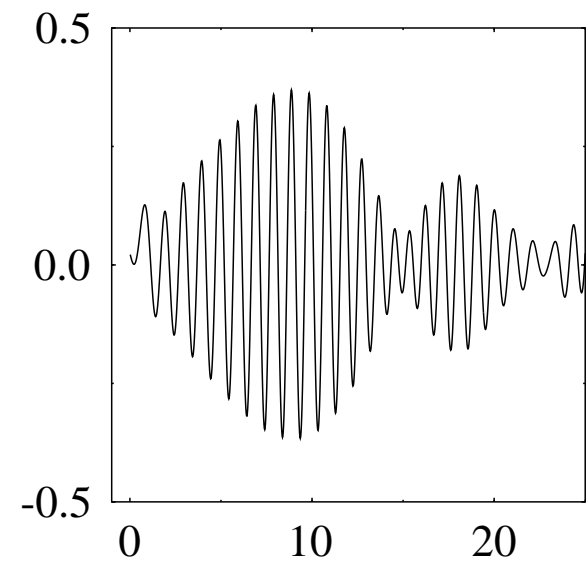
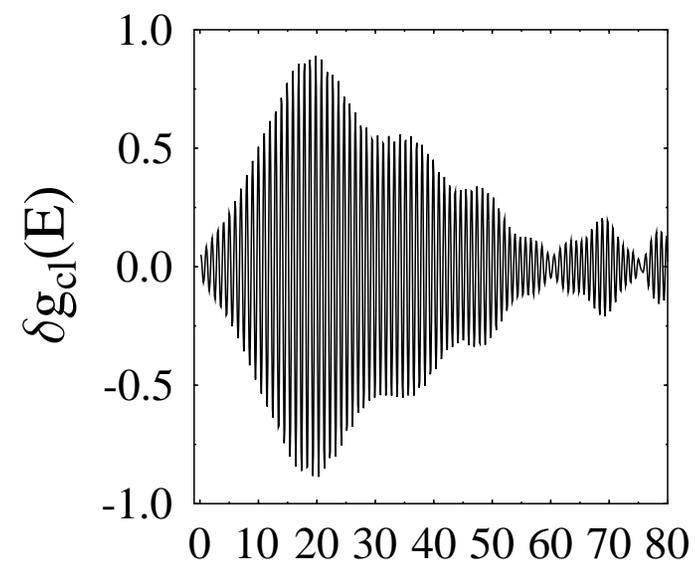 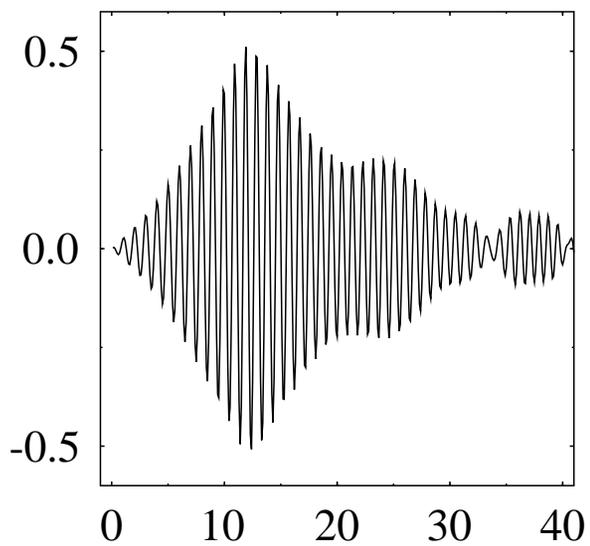 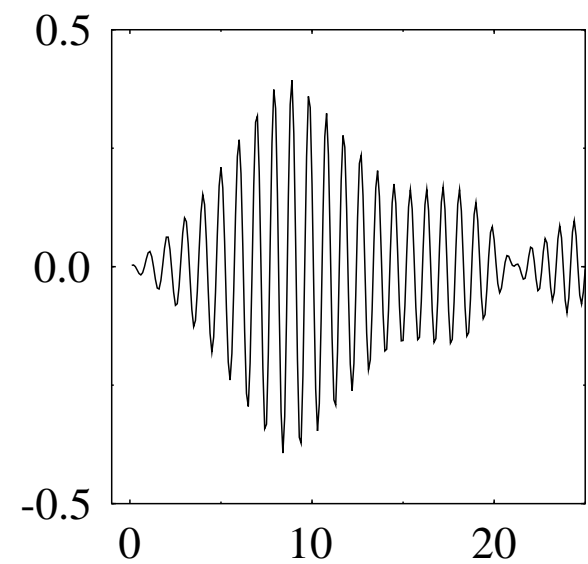

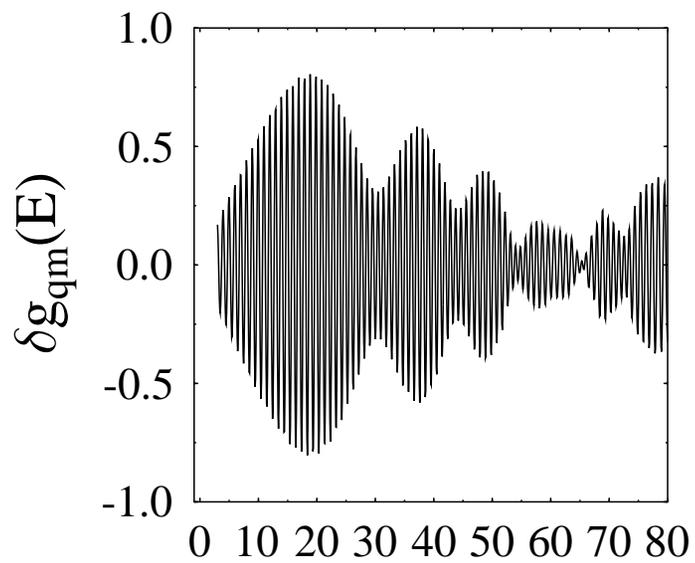 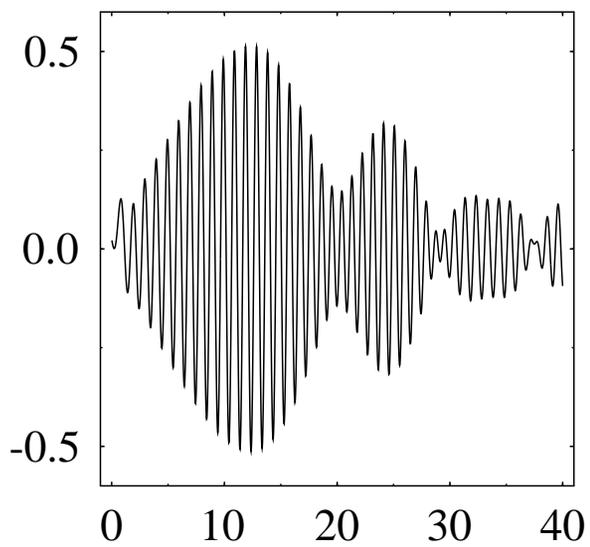 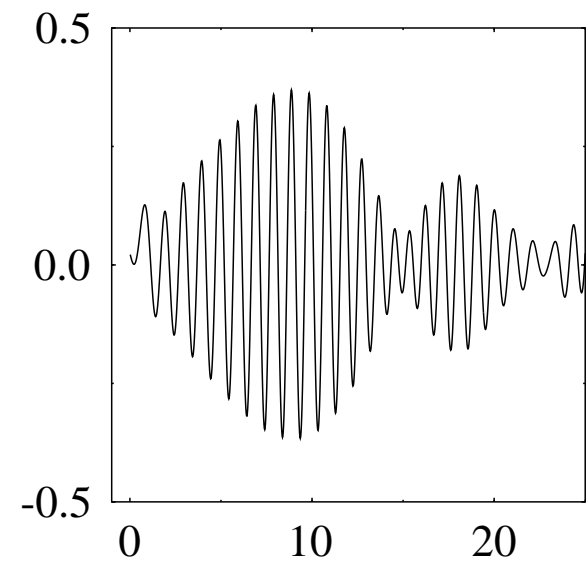
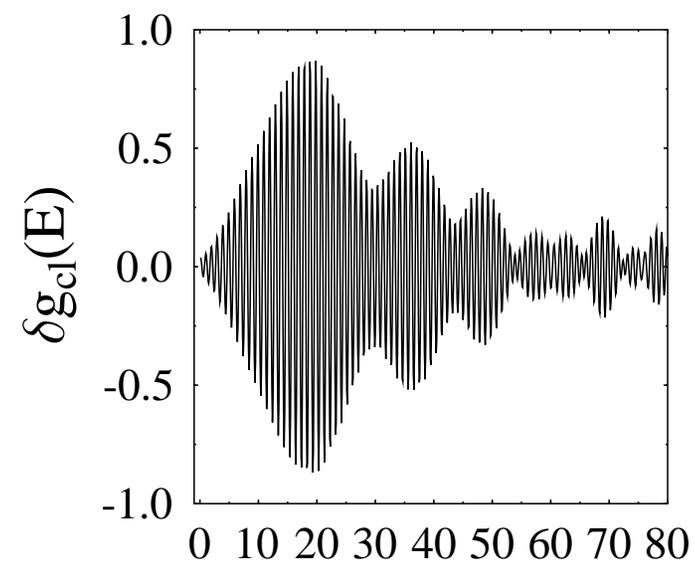 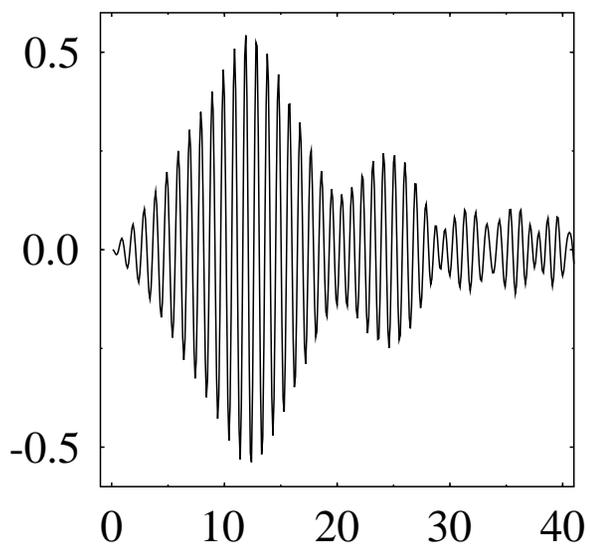 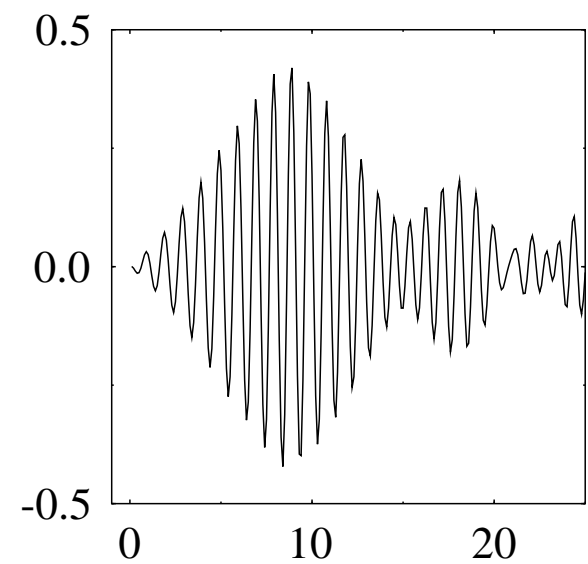

E (α=0.04)  E (α=0.06)  E (α=0.08)

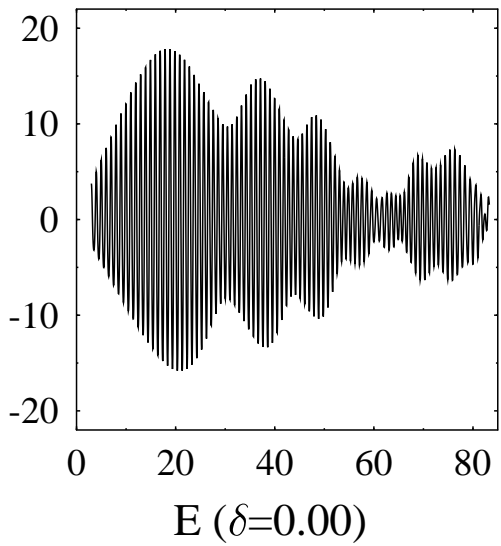
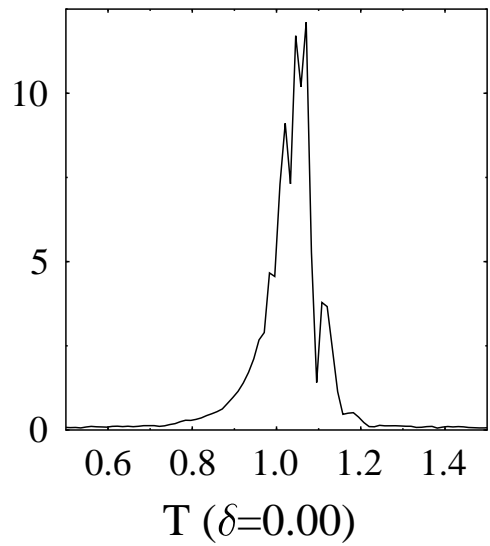
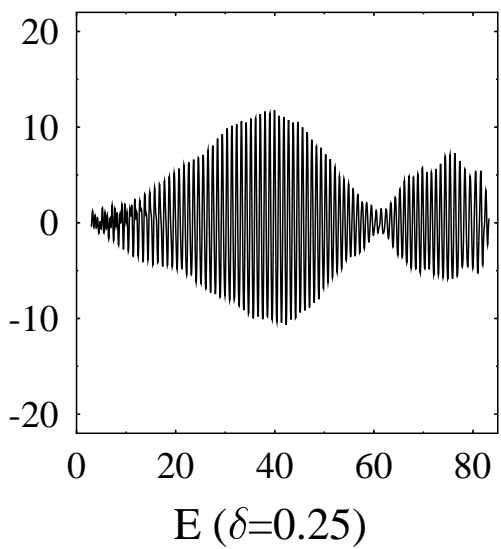
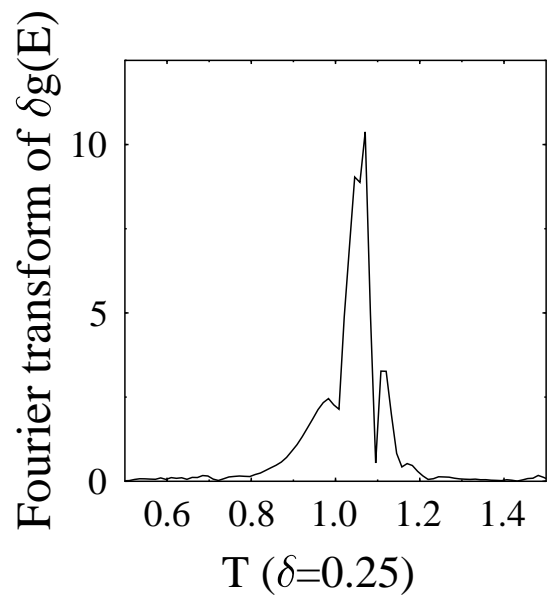
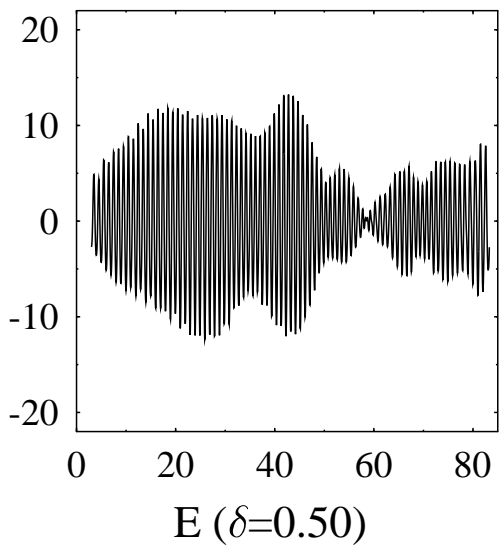
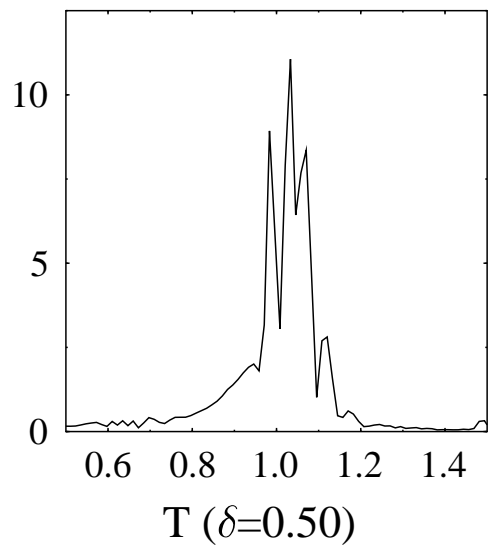

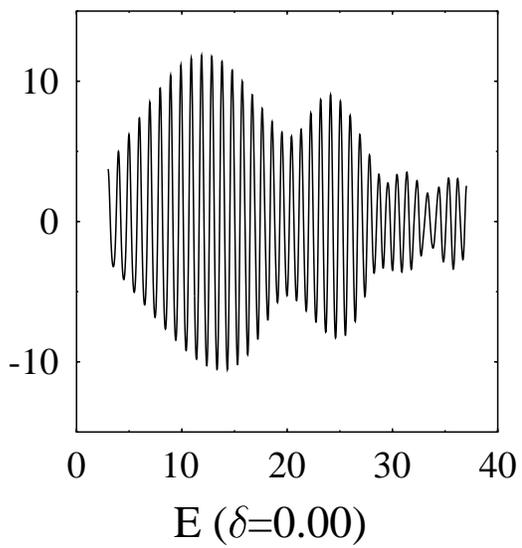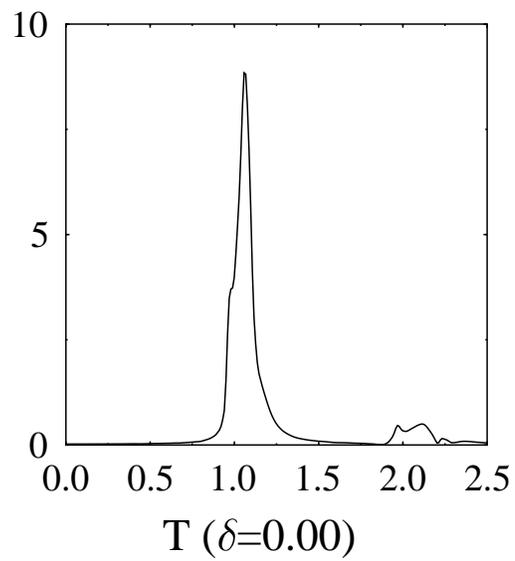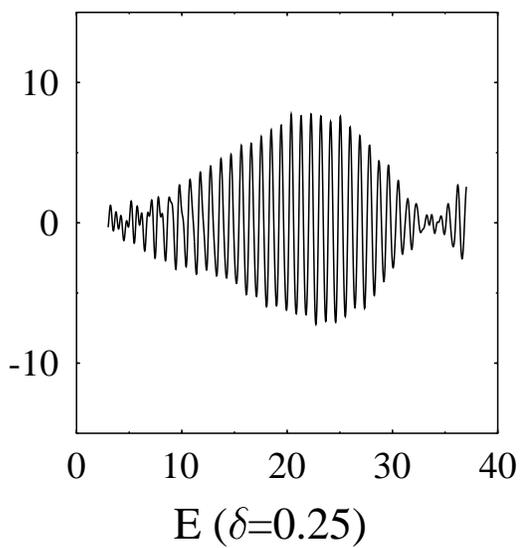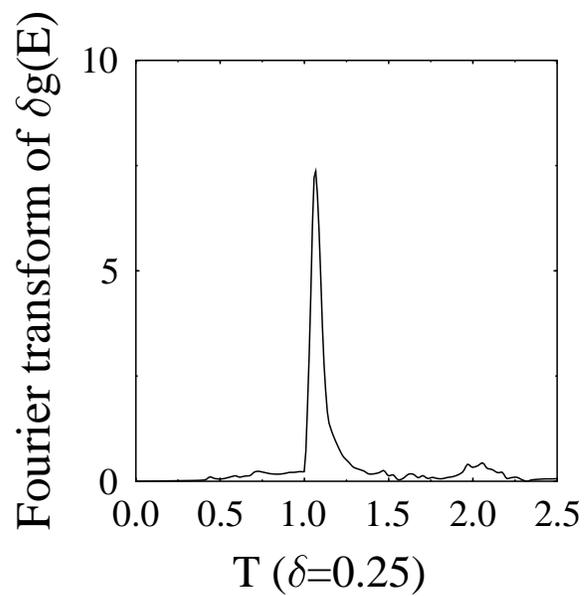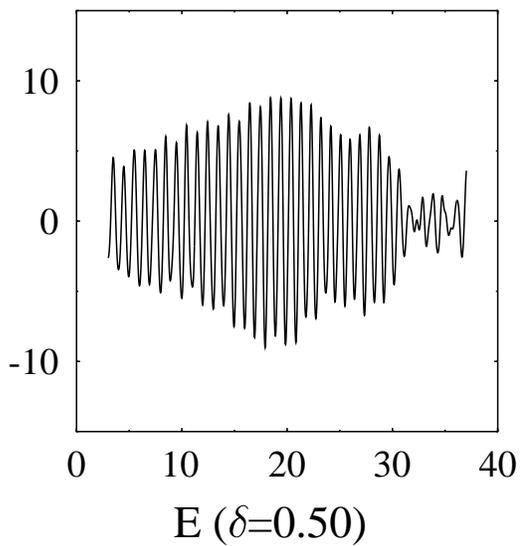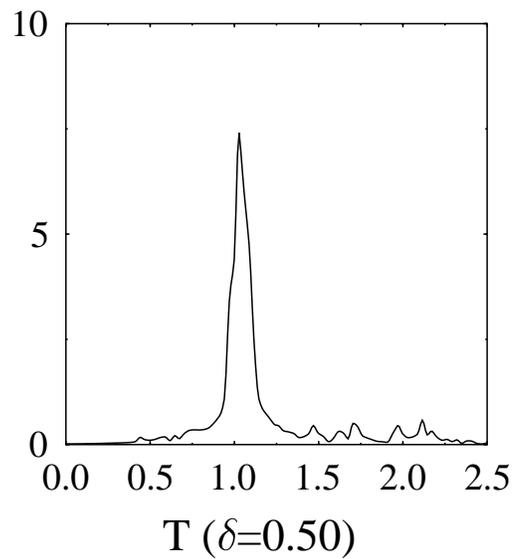

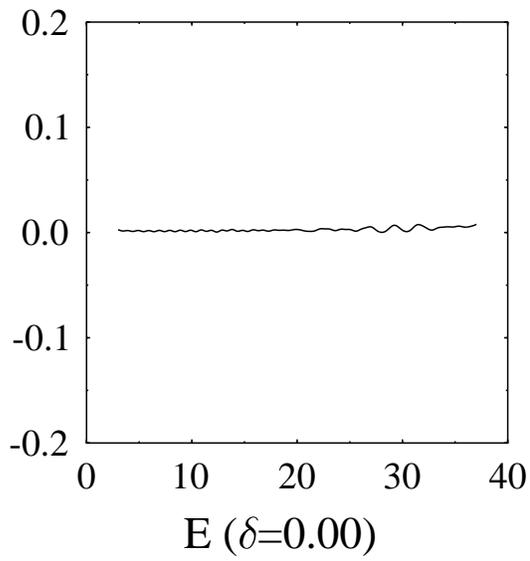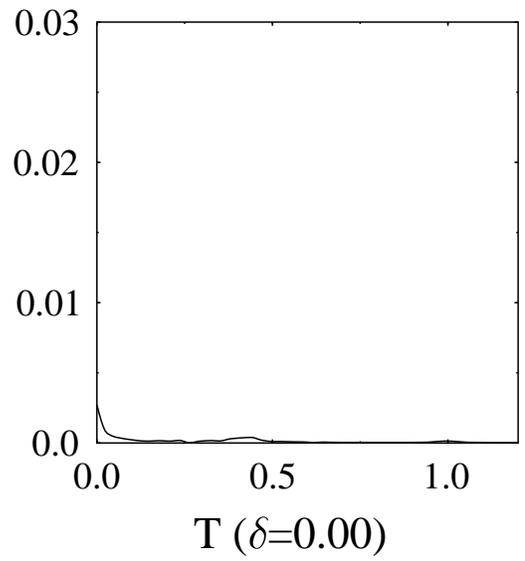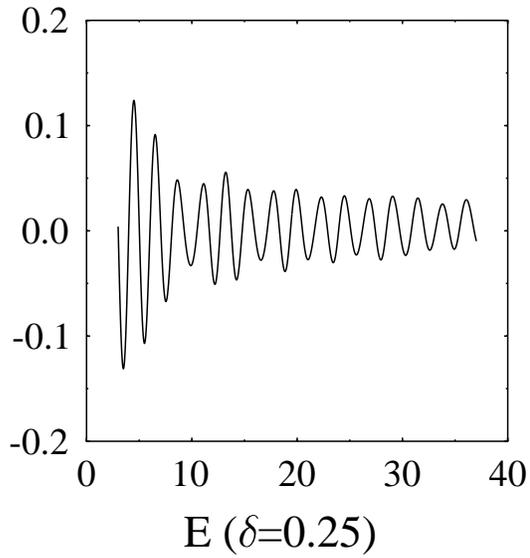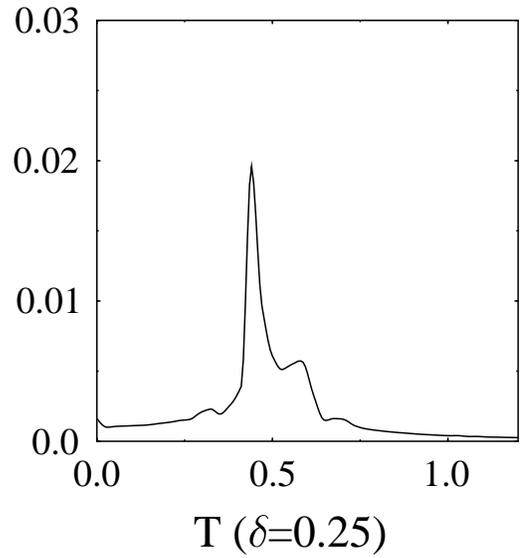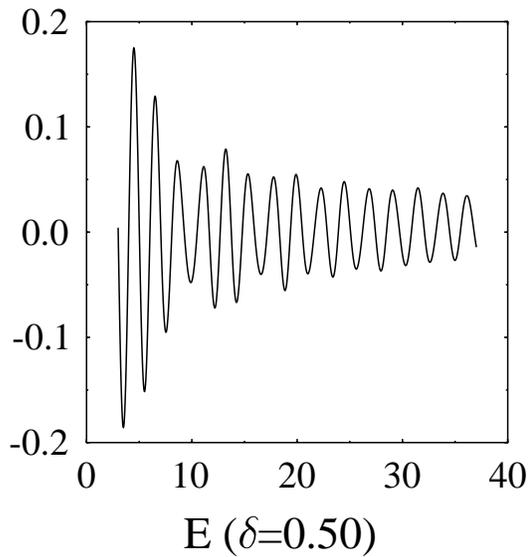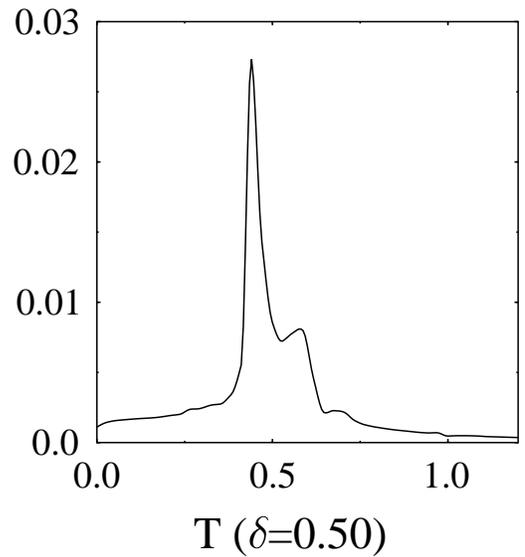

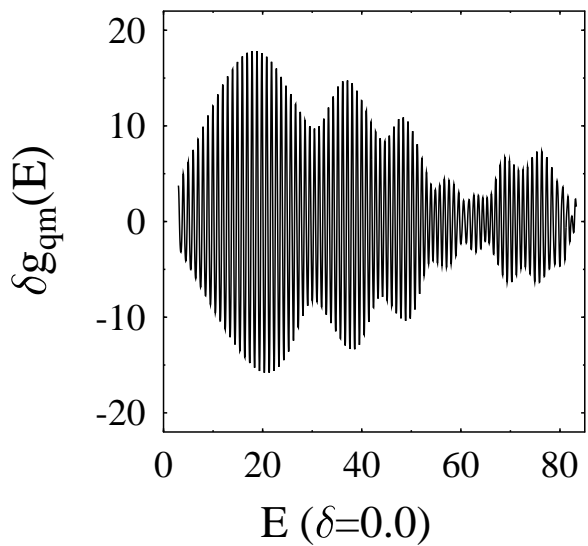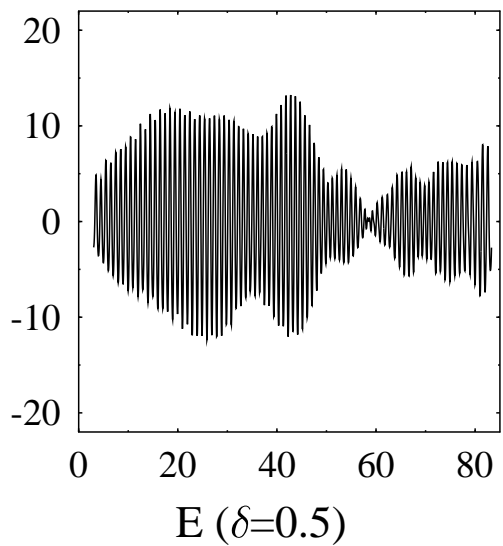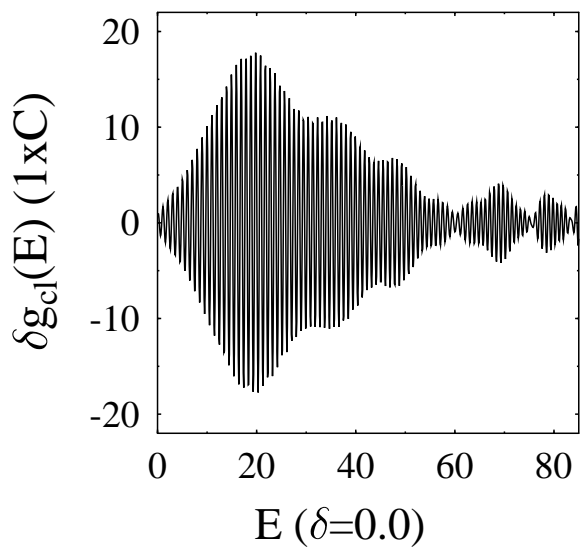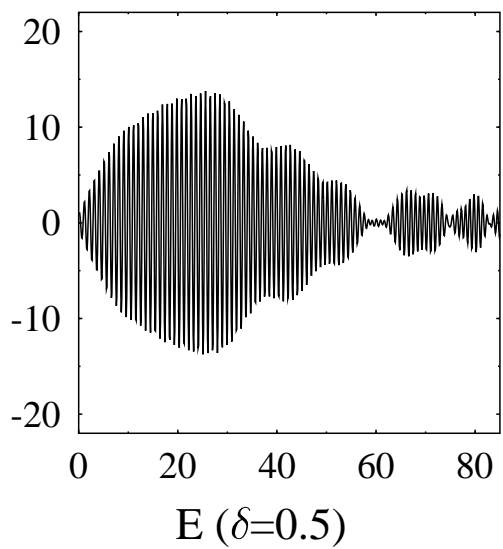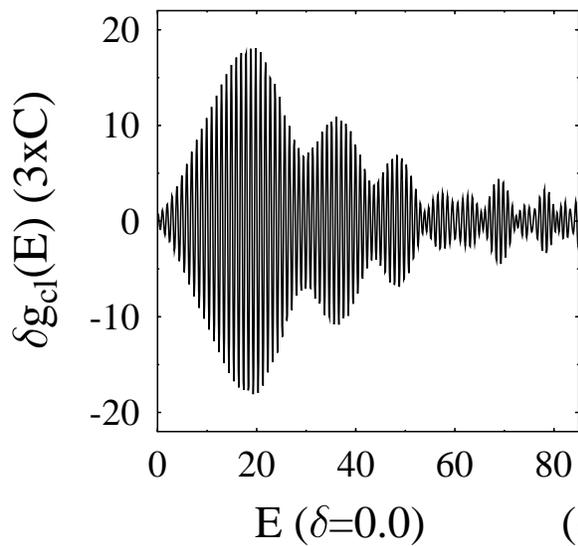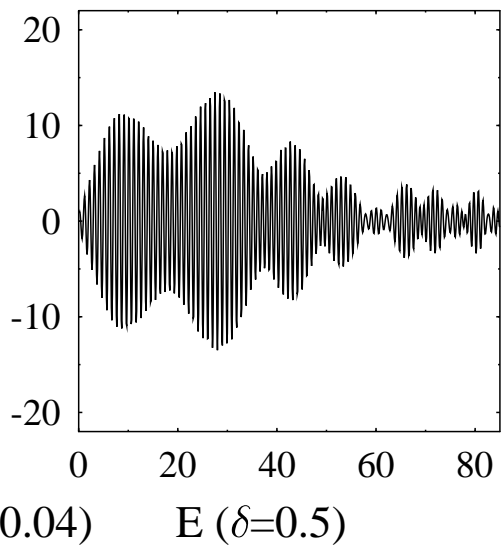

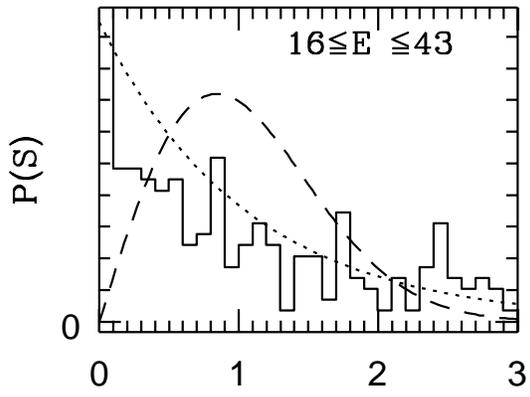
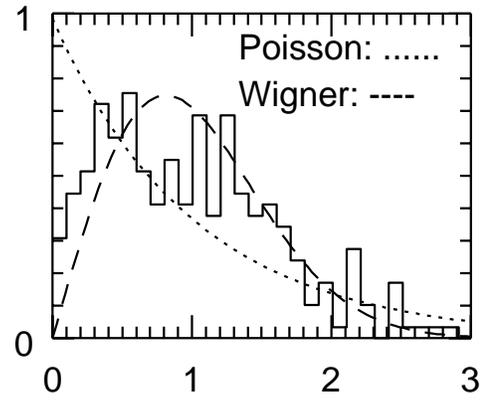
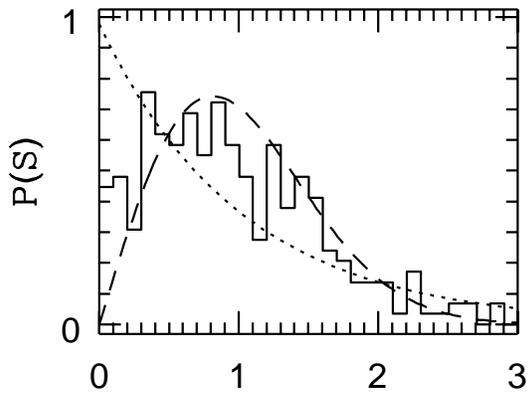
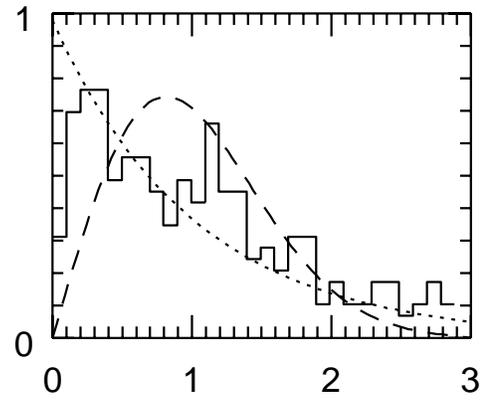
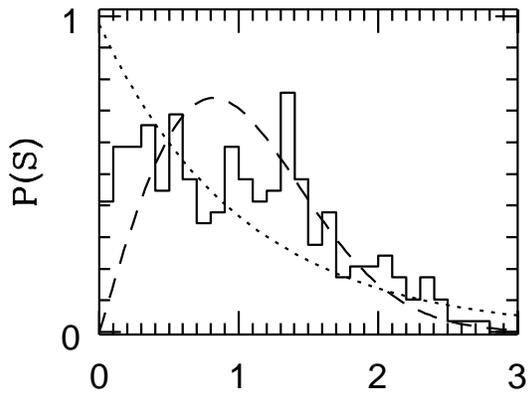
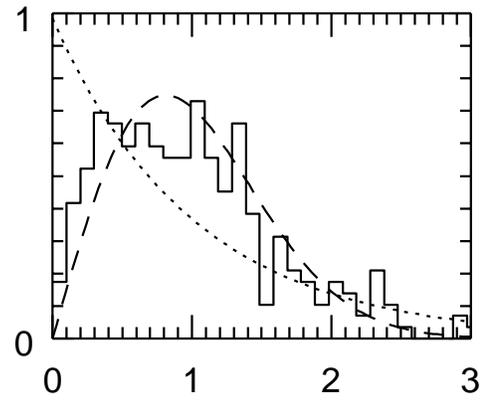

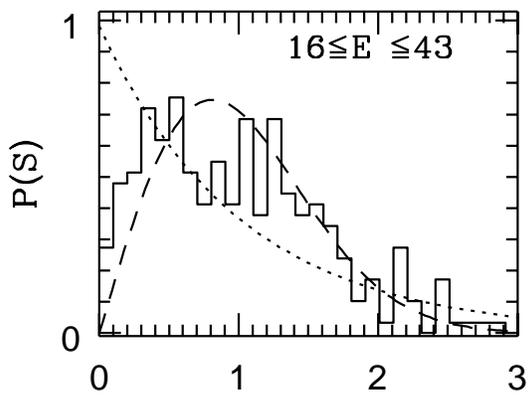
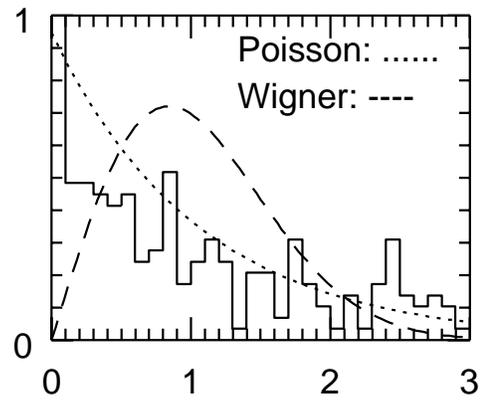
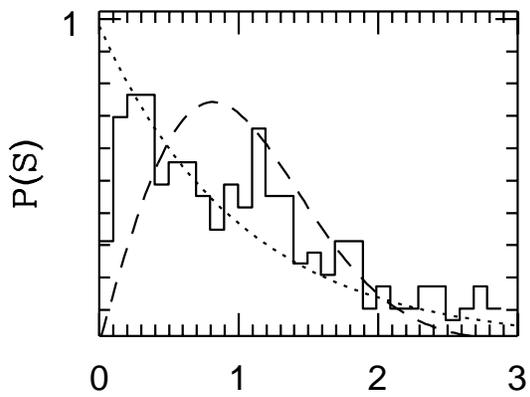
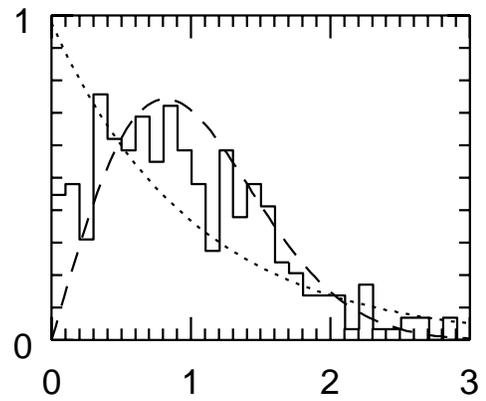
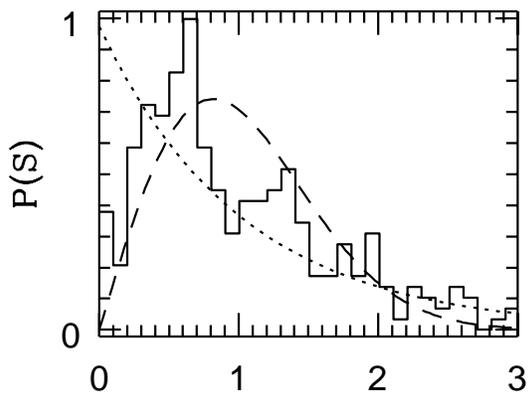
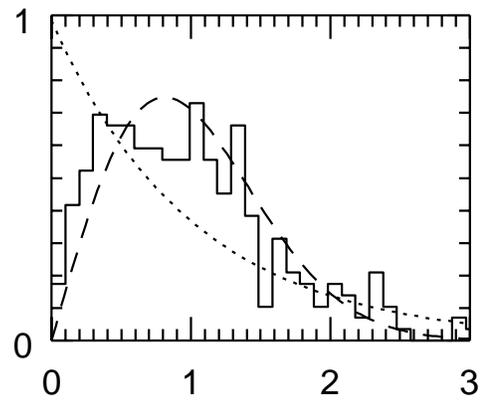

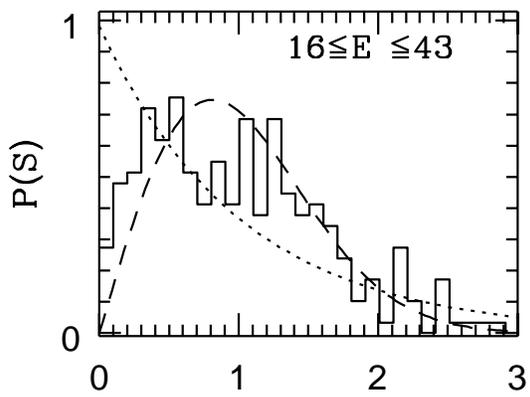
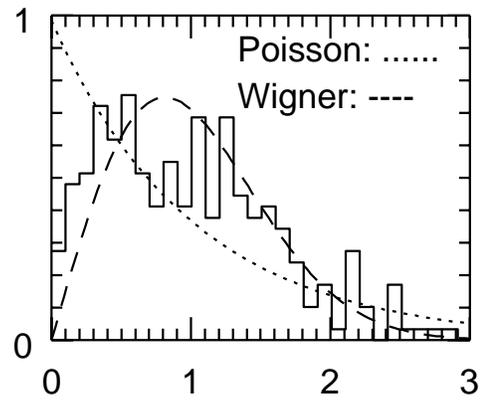
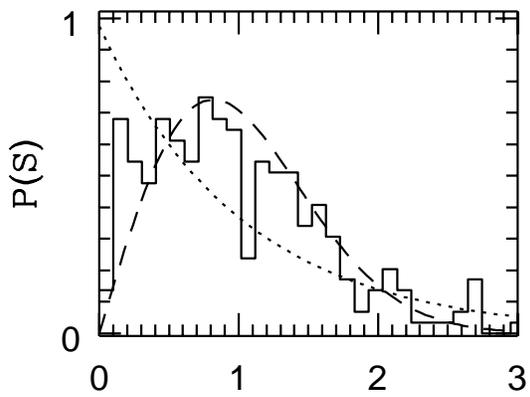
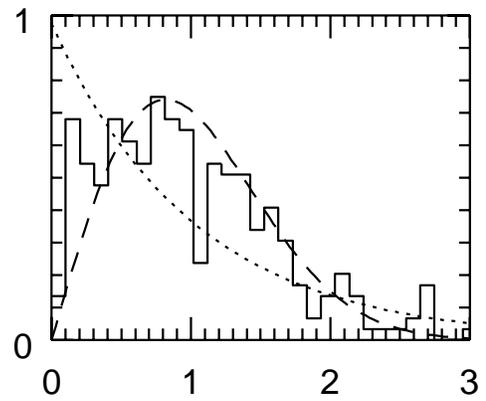
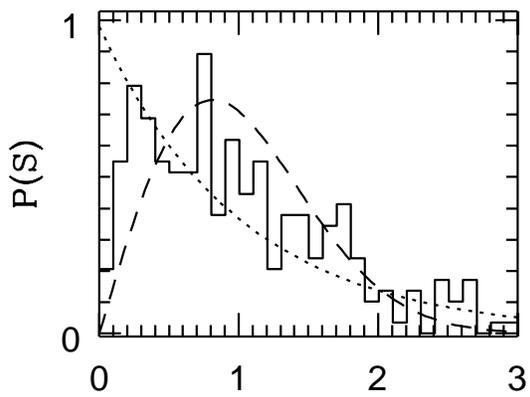
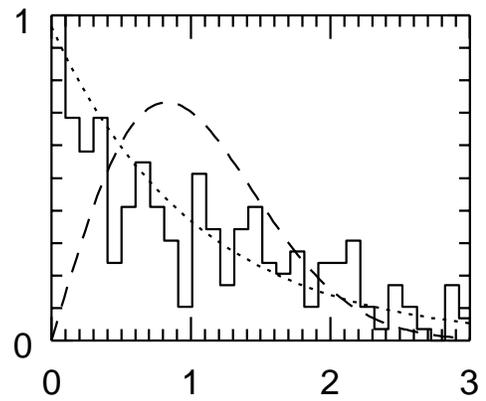

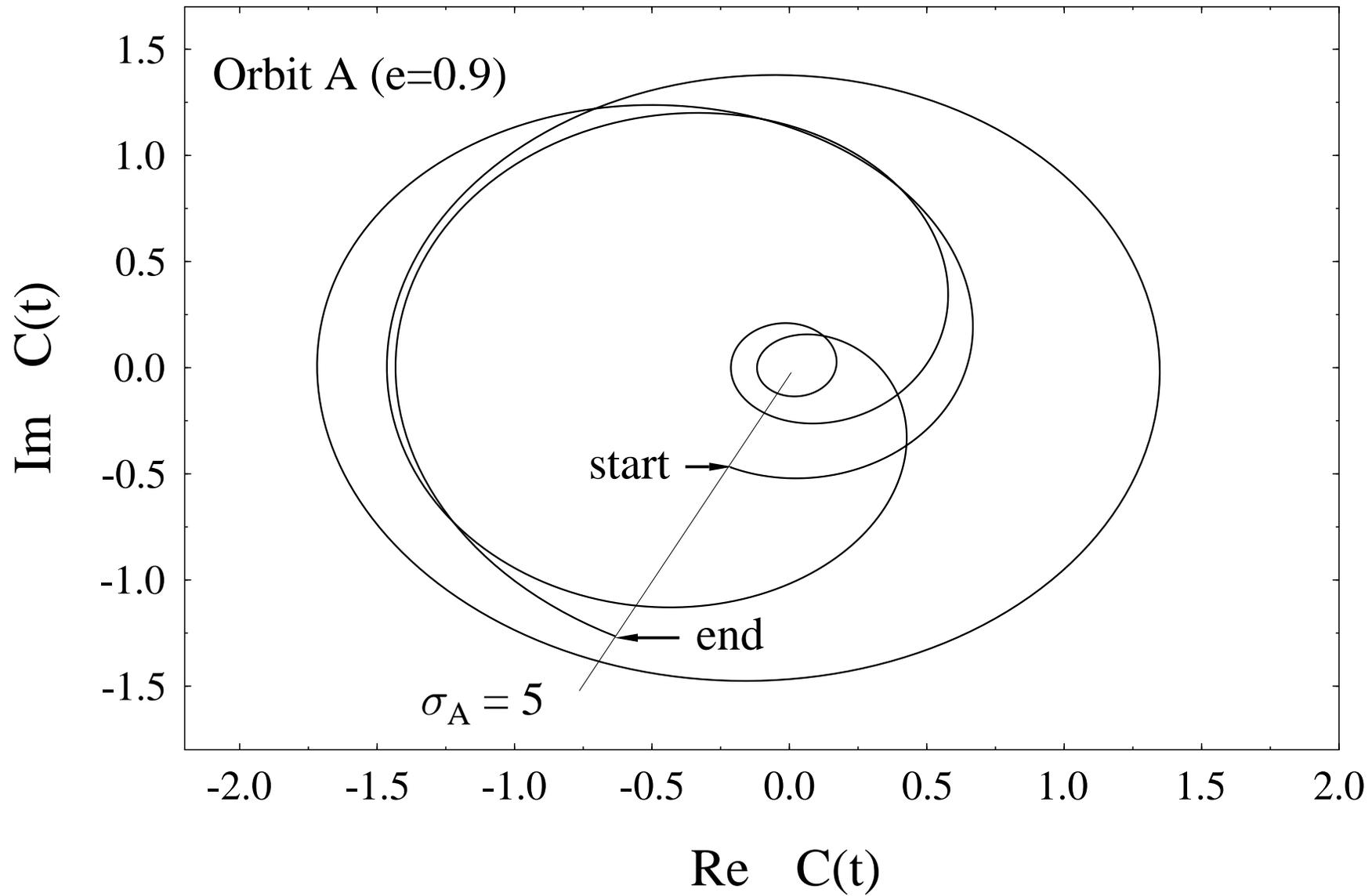

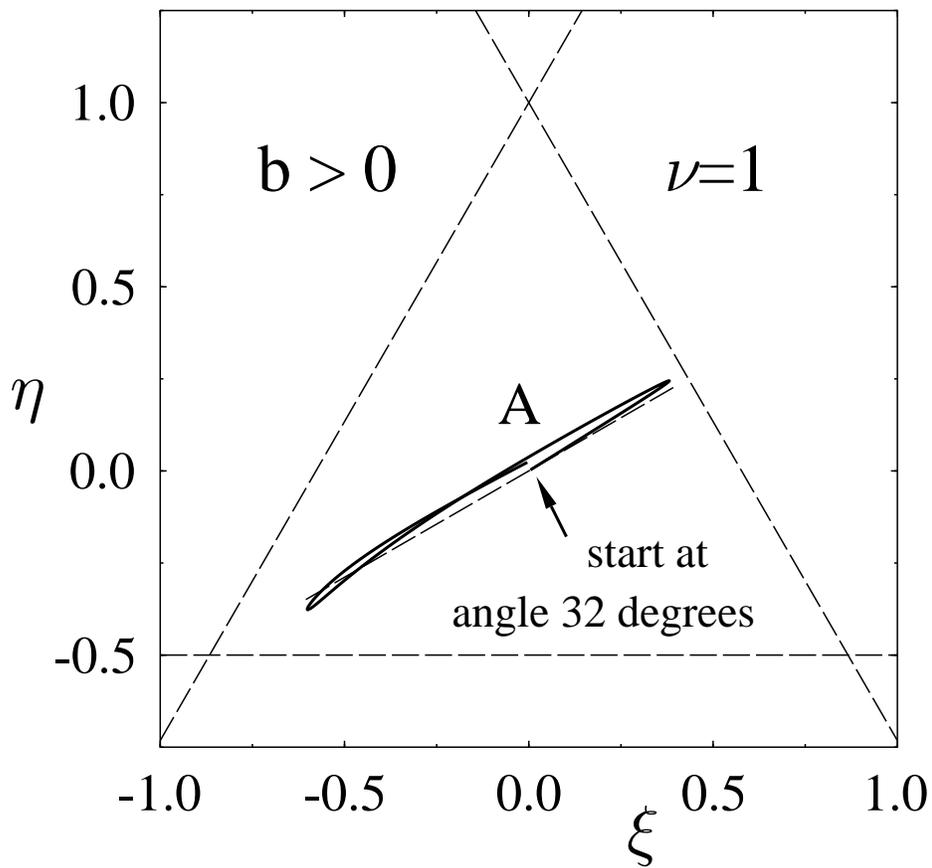
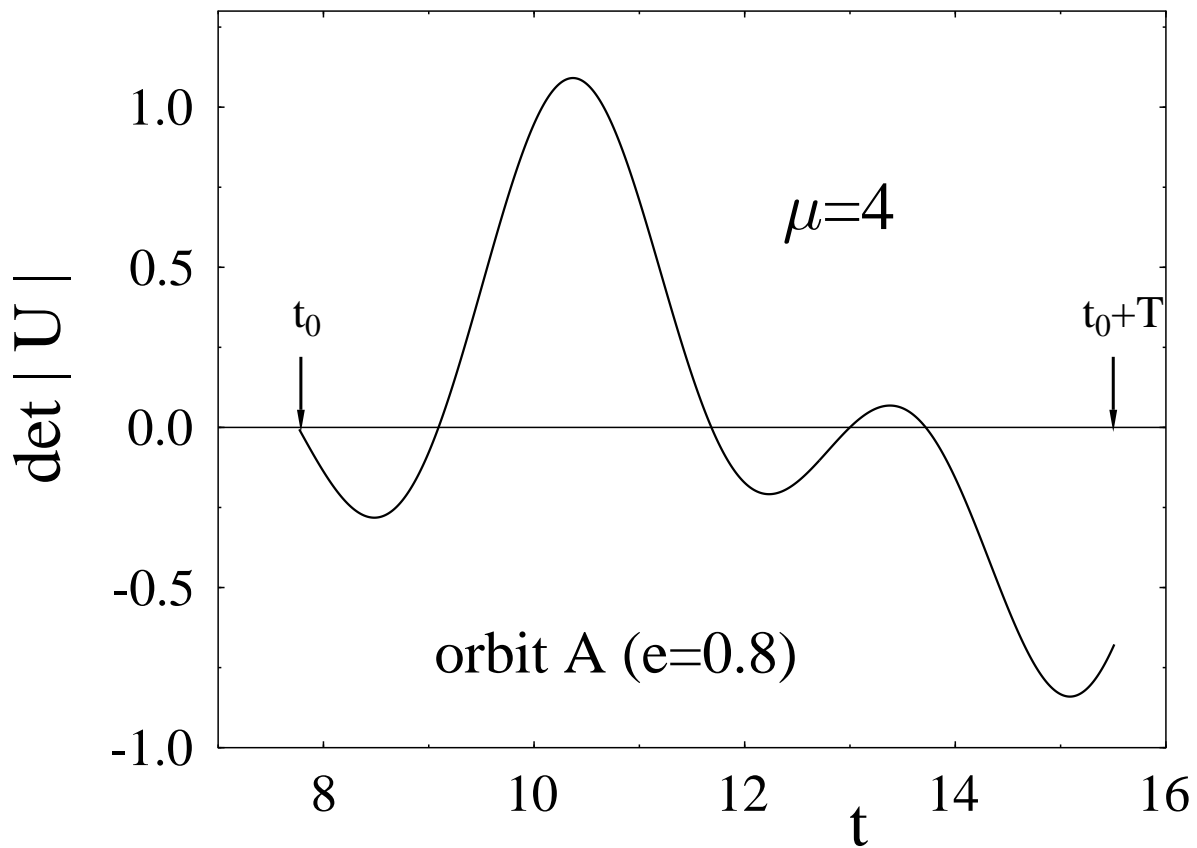